\documentclass[newfonts=false,format=sigconf,10pt,letterpaper]{acmart}
\renewcommand\footnotetextcopyrightpermission[1]{} 

\makeatletter
\renewcommand\@formatdoi[1]{\ignorespaces}
\makeatother

\setcopyright{none}

\acmDOI{10.475/123_4}

\acmISBN{123-4567-24-567/08/06}

\acmConference[SHORTNAME'17]{ACM Long Conference Name conference}{July 1997}{City, State, Country} 
\acmYear{2017}
\copyrightyear{2017}

\acmPrice{15.00}

\usepackage{booktabs} 
\usepackage{subcaption}

\usepackage{graphicx, multirow, times}

\usepackage{xurl}
\usepackage{amsfonts, verbatim, mathtools, color, xspace, caption, amsmath}
\usepackage{listings}
\usepackage{color}
\usepackage{xcolor}
\usepackage{array}
\usepackage{colortbl}
\usepackage{alltt}
\usepackage[small,compact]{titlesec}
\usepackage[labelfont=bf]{caption}

\usepackage{mdwlist}

\newcommand{\eg}{{\it e.g.,}\xspace}
\newcommand{\ie}{{\it i.e.,}\xspace}

\usepackage{natbib}

\newcommand{\mypara}[1]{\vspace{0.07cm}\noindent{\bf {#1}:}~}

\usepackage{listings}
\usepackage{subcaption}
\usepackage{upgreek}
\usepackage[ruled,vlined,linesnumbered]{algorithm2e}

\usepackage[normalem]{ulem}

\SetKw{KwBy}{by}
\SetKwInOut{Input}{Input}

\SetCommentSty{mycommfont}

\newlength\mylen

\newcommand{\eat}[1]{}
\newcommand{\name}{{\tt Ekya}\xspace}

\newcommand{\fair}{uniform\xspace}
\newcommand{\Fair}{Uniform\xspace}

\newcommand{\junchen}[1]{{\color{blue} [Junchen: {#1}]}}

\newcounter{packednmbr}

\newenvironment{packeditemize}{\begin{list}{$\bullet$}{\setlength{\itemsep}{0.5pt}\addtolength{\labelwidth}{-4pt}\setlength{\leftmargin}{2ex}\setlength{\listparindent}{\parindent}\setlength{\parsep}{1pt}\setlength{\topsep}{0pt}}}{\end{list}}

\title{{\huge Ekya: Continuous Learning of Video Analytics Models on Edge Compute Servers}}
\author{\Large Romil Bhardwaj}
\affiliation{\institution{\normalsize Microsoft, UC Berkeley}}
\author{\Large Zhengxu Xia}
\affiliation{\institution{University of Chicago}}
\author{\Large Ganesh Ananthanarayanan}
\affiliation{\institution{Microsoft}}
\author{\Large Junchen Jiang}
\affiliation{\institution{University of Chicago}}
\author{\Large Nikolaos Karianakis}
\affiliation{\institution{Microsoft}}
\author{\Large Yuanchao Shu}
\affiliation{\institution{Microsoft}}
\author{\Large Kevin Hsieh}
\affiliation{\institution{Microsoft}}
\author{\Large Victor Bahl}
\affiliation{\institution{Microsoft}}
\author{\Large Ion Stoica}
\affiliation{\institution{UC Berkeley}}

\date{}

\begin{document}

        
\maketitle
\section*{Abstract}
Video analytics applications use edge compute servers for the analytics of the videos (for bandwidth and privacy).  
Compressed models that are deployed on the edge servers for inference suffer from {\em data drift} where the live video data diverges from the training data. Continuous learning handles data drift by periodically retraining the models on new data. Our work addresses the challenge of jointly supporting inference and retraining tasks on edge servers, which requires navigating the fundamental tradeoff between  the retrained model's accuracy and
the inference accuracy. 
Our solution \name balances this tradeoff across multiple models and uses a micro-profiler to identify the models that will benefit the most by retraining. {\name}'s accuracy gain compared to a baseline scheduler is $29\%$ higher, and the baseline requires $4\times$ more GPU resources to achieve the same accuracy as \name.


\section{Introduction}
\label{sec:intro}

Video analytics applications, such as for urban mobility and smart cars \cite{bellevue-report}, are being powered by deep neural network (DNN) models 
for object detection and classification, e.g., Yolo \cite{yolo9000-1}, ResNet \cite{deepresidual-2} and EfficientNet \cite{efficientnet-3}. 
Video analytics deployments stream the videos to {\em edge servers} \cite{azure-ase, aws-outposts} placed on-premise \cite{ieee-computer, edgevideo-1, edgevideo-2, getmobile}. Edge computation is preferred for video analytics as it does not require expensive network links to stream videos to the cloud \cite{getmobile}, while also ensuring privacy of the videos (e.g., many European cities mandate against streaming their videos to the cloud \cite{sweden-data, azure-data}).

Edge compute is provisioned with limited resources (e.g., with weak GPUs \cite{aws-outposts, azure-ase}). This limitation is worsened by the mismatch between the growth rate of the compute demands of models and the compute cycles of processors \cite{ion-blog, openai-blog}. As a result, edge deployments rely on {\em model compression} \cite{compression-4, compression-5, compression-6}. 
The compressed DNNs are initially trained on representative data from each video stream, but while in the field, they are affected by {\em data drift}, i.e., the live video data diverges significantly from the data that was used for training \cite{datadrift-7, datadrift-8, datadrift-a, datadrift-b}. Cameras in streets and smart cars encounter  
varying scenes over time, e.g., lighting, crowd densities, 
and changing object mixes. 
It is difficult to exhaustively cover all these variations in the training, especially since even subtle variations affect the accuracy. As a result, there is a sizable drop in the accuracy of edge DNNs due to data drift (by 22\%; \S\ref{subsec:continuous-measurement}). 
In fact, 
the fewer weights and shallower architectures of compressed DNNs often make them unsuited to provide high accuracy when trained with large variations in the data.

\noindent{\bf Continuous model retraining.} A promising approach to address data drift is continuous learning. The edge DNNs are incrementally {\em retrained} on the new video samples even as some knowledge from before is retained \cite{compressiondrift-11, continuous-12}. Continuous learning techniques retrain the DNNs periodically 
\cite{distribution-20, mullapudi2019}; we refer to the period between two retrainings as the ``retraining window'' and we use a sample of the data that is accumulated during each window for retraining. 
Such ongoing learning \cite{incremental-13, icarl-14, incremental-15} 
helps the compressed edge models maintain high accuracy even with changing data characteristics.

Edge servers use their GPUs \cite{azure-ase} for DNN inference on many live video streams (e.g., traffic cameras in a city).  
Adding continuous training to edge servers presents a tradeoff between the live inference accuracy and drop in accuracy due to data drift. Allocating more resources to the retraining job allows it to finish faster and provide a more accurate model sooner. At the same time, during the retraining, taking away resources from the inference job lowers its accuracy (because it may have to sample the frames of the video to be analyzed).


Central to the resource demand and accuracy of the jobs 
are their {\em configurations}. For retraining jobs, configurations refer to the hyperparameters, e.g., number of training epochs, that substantially impact the resource demand and accuracies (\S\ref{subsec:profiles}). The improvement in accuracy due to retraining also depends on {\em how much} the characteristics of the live videos have changed. For inference jobs, configurations like frame sampling and resolution impact the accuracy and resources needed to keep up with analyzing the live video \cite{chameleon, noscope}. 
\noindent{\bf Problem statement.} 
We make the following decisions for continuous retraining. 
($1$) in each retraining window, decide which of the edge models to retrain;  
($2$) allocate the edge server's GPU resources among the retraining and inference jobs, and 
($3$) select the configurations of the retraining and inference jobs. 
We also constraint our decisions such that the inference accuracy {\em at any point in time} does not drop below a minimum value (so that the outputs continue to remain useful to the application). 
Our objective in making the above three decisions is to maximize the inference accuracy {\em averaged over the retraining window} (aggregating the accuracies during and after the retrainings) across all the videos analyzed on the edge server. 
Maximizing the inference accuracy over the retraining window creates new challenges as it is different from $(i)$ video inference systems that optimize only the instantaneous accuracy \cite{videostorm, noscope, chameleon}, $(ii)$  model training systems that optimize only the eventual accuracy \cite{hyperparameter-16, DBLP:journals/jmlr/BergstraB12, DBLP:conf/nips/SnoekLA12, Swersky_scalablebayesian, DBLP:conf/osdi/XiaoBRSKHPPZZYZ18, DBLP:conf/eurosys/PengBCWG18}.

Addressing the fundamental tradeoff between the retrained model's accuracy and the inference accuracy is computationally complex. 
First, the decision space is multi-dimensional consisting of a diverse set of retraining and inference configurations, and choices of resource allocations over time. 
Second, it is difficult to know the performance of different configurations (in resource usage and accuracy) as it requires actually retraining using different configurations. Data drift exacerbates these challenges because a decision that works well in a retraining window may not do so in the future.
\noindent{\bf Solution components.} Our solution {\name} has two main components: a resource scheduler and a performance estimator. 

In each retraining window, the {\bf resource scheduler} makes the three decisions listed above in our problem statement. 
In its decisions, {\name}'s scheduler prioritizes retraining the models of those video streams whose characteristics have changed the most because these models have been most affected by data drift. 
The scheduler decides against retraining the models which do not improve our target metric.  
To prune the large decision space, the scheduler uses the following techniques. First, it simplifies the spatial complexity by considering GPU allocations only in coarse fractions (e.g., 10\%) that are  accurate enough for the scheduling decisions,  while also being mindful of the granularity achievable in modern GPUs \cite{nvidia-mps}. 
Second, it does not change allocations to jobs {\em during the retraining}, thus largely sidestepping the temporal complexity. Finally, our micro-profiler (described below) prunes the list of configurations to only the promising options.

To make efficient choices of configurations, the resource scheduler relies on estimates of accuracy after the retraining and the resource demands. We have designed a {\bf micro-profiler} that observes the accuracy of the retraining configurations on a {\em small subset} of the training data in the retraining window with {\em just a few epochs}. It uses these observations to extrapolate the accuracies when retrained on a larger dataset for many more epochs. Further, we restrict the micro-profiling to only a small set of {\em promising} retraining configurations. Together, these techniques result in {\name}'s micro-profiler being nearly $100\times$ more efficient than exhaustive profiling while still estimating accuracies with an error of $5.8\%$. To estimate the resource demands, the micro-profiler measures the retraining duration {\em per epoch} when $100\%$ of the GPU is allocated, and scales out the training time for different allocations, number of epochs, and training data sizes. 

\noindent{\bf Implementation and Evaluation.} We have implemented and evaluated {\name} using a system deployment and trace-driven simulation. We used video workloads from dashboard cameras of smart cars (Waymo \cite{waymo} and Cityscapes \cite{cityscapes}) as well as from statically mounted traffic and building cameras over 24 hour durations. 
{\name}'s accuracy compared to competing baselines is $29\%$. higher. As a measure of {\name}'s efficiency, attaining the same accuracy as {\name} will require $4\times$ more GPU resources on the edge server with the baseline. 

\noindent{\bf Contributions:} Our work makes the following contributions. 


\noindent{1)} We introduce the metric of {\em inference accuracy averaged over the retraining window} for continuous training systems.

\noindent{2)} We design an {\em efficient micro-profiler to estimate} the benefits and costs of retraining edge DNN models. 

\noindent{3)} We design a scalable resource scheduler for {\em joint retraining and inference} on edge servers. 
 
\noindent{\bf This work does not raise any ethical issues.}
\section{Continuous training of models on edge compute}
\label{sec:background}


\subsection{Edge Computing for Video Analytics}
\label{subsec:edge}

\begin{figure}[t]
    \centering
    \includegraphics[width=0.7\columnwidth]{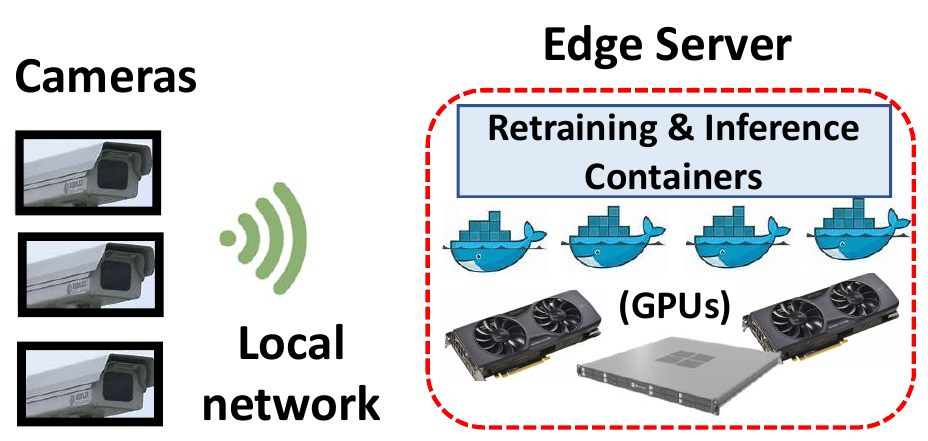}
    \caption{\bf\small Cameras connect to the edge server (e.g., Azure Stack Edge \cite {azure-ase}) through local networks (wireless, typically). The edge server is equipped with consumer-grade GPUs and executes DNN inference containers and retraining containers.
    }
    \label{fig:edge}
\end{figure}

Video analytics deployments commonly analyze videos on edge servers placed on-premise (e.g., AWS Outposts \cite{aws-outposts} or Azure Stack Edge \cite{azure-ase}). 
Due to cost and energy constraints, compute efficiency is one of the key design goals of edge computing. 
A typical edge server supports tens of video streams \cite{videoedge}, e.g., on the cameras in a building, with customized analytics and models for each stream \cite{rocket-github} (see Figure \ref{fig:edge}).

Video analytics applications adopt edge computing for reasons of limited network bandwidth to the cloud, unreliability of the network, and privacy of the video content \cite{getmobile, edgevideo-1, ieee-computer}. 

$1)$ Edge deployments are often in locations where the {\em uplink network to the cloud is expensive} for shipping continuous video streams, 
e.g., in oil rigs with expensive satellite network or smart cars with data-limited cellular network. 
\footnote{The uplinks of LTE cellular or satellite links is $3-10$Mb/s \cite{39-getmobile, 57-getmobile}, which can only support a couple of $1080$p 30 fps HD video streams whereas a typical deployment has many more cameras \cite{getmobile}.}

$2)$ Network links out of the edge locations experience {\em outages} \cite{20-getmobile, getmobile}. 
Edge compute provides robustness against disconnection to the cloud \cite{chick-fill} and prevents disruptions \cite{37-getmobile}. 

$3)$ Videos often contain {\em sensitive and private data} that users do not want sent to the cloud (e.g., many EU cities legally mandate that traffic videos be processed on-premise \cite{sweden-data, azure-data}). 

Thus, due to reasons of network cost and video privacy, it is preferred to run both inference and retraining on the edge compute device itself without relying on the cloud. In fact, with bandwidths typical in edge deployments, cloud-based solutions are slower and result in lower accuracies (\S\ref{subsec:eval-alternate}).

\subsection{Compressed DNN Models and Data drift}
\label{subsec:continuous}

Advances in computer vision research have led to high-accuracy DNN models that 
achieve high accuracy with a large number of weights, deep architectures, and copious training data. While highly accurate, using these heavy and general DNNs for video analytics is both expensive and slow \cite{noscope, DBLP:conf/osdi/HsiehABVBPGM18}, which make them unfit for resource-constrained edge computing. The most common approach to addressing the resource constraints on the edge is to train and deploy \emph{specialized and compressed} DNNs \cite{compression-4, compression-5, compression-6, compression-17, compression-18, compression-19}, which consist of far fewer weights and shallower architectures. These compressed DNNs are trained to only recognize the limited objects and scenes specific to each video stream. In other words, to maintain high accuracy, they forego generality for improved compute efficiency \cite{noscope, DBLP:conf/osdi/HsiehABVBPGM18, mullapudi2019}.

\noindent{\bf Data drift:} As specialized edge DNNs have fewer weights and shallower architectures than general DNNs, they can only memorize limited amount of object appearances, object classes, and scenes. As a result, specialized edge DNNs are particularly vulnerable to {\em data drift} \cite{datadrift-7, datadrift-8, datadrift-a, datadrift-b}, where live video data diverges significantly from the initial training data. For example, variations in the angles of objects, scene density (e.g. rush hours), and lighting (e.g., sunny vs. rainy days) over time make it difficult for traffic cameras to accurately identify the objects of interest (cars, bicycles, road signs). Cameras in modern cars observe vastly varying scenes (e.g., building colors, crowd sizes) as they move through different neighborhoods and cities. Further, the {\em distribution} of the objects change over time, which in turn, reduces the edge model's accuracy \cite{distribution-20, distribution-21}. Owing to their ability to memorize limited amount of object variations, to maintain high accuracy, edge DNNs have to be continuously updated with the recent data and to the changing object distributions.  

\noindent{\bf Continuous training:} The preferred approach, that has gained significant attention, is for edge DNNs to {\em continuously learn} as they incrementally observe new samples over time \cite{incremental-13, icarl-14, incremental-15}. 
The high temporal locality of videos allows the edge DNNs to focus their learning on the most recent object appearances and object classes \cite{DBLP:conf/cvpr/ShenHPK17, mullapudi2019}.  
In \name, we use a modified version of iCaRL \cite{icarl-14} though our techniques are generally applicable. Our learning algorithm on-boards new classes, as well as adapts to the changing characteristics of the existing classes. 
Since manual labeling is not feasible for continuous training systems on the edge, we obtain the labels using a ``golden model'' that is highly accurate but is far more expensive because it uses a deeper architecture with large number of weights. The golden model cannot keep up with inference on the live videos and we use it to label only a small fraction of the videos in the retraining window that we use for retraining. Our approach is essentially that of supervising a low-cost ``student'' model with a high-cost ``teacher'' model (or knowledge distillation \cite{44873}), and this has been broadly applied in computer vision literature \cite{incremental-13, mullapudi2019, incremental-15, distribution-20}. 

\subsection{Accuracy benefits of continuous learning}
\label{subsec:continuous-measurement}

\begin{figure}[t!]
  \centering
  \begin{subfigure}[t]{0.5\linewidth}
    \centering
    \includegraphics[width=\linewidth]{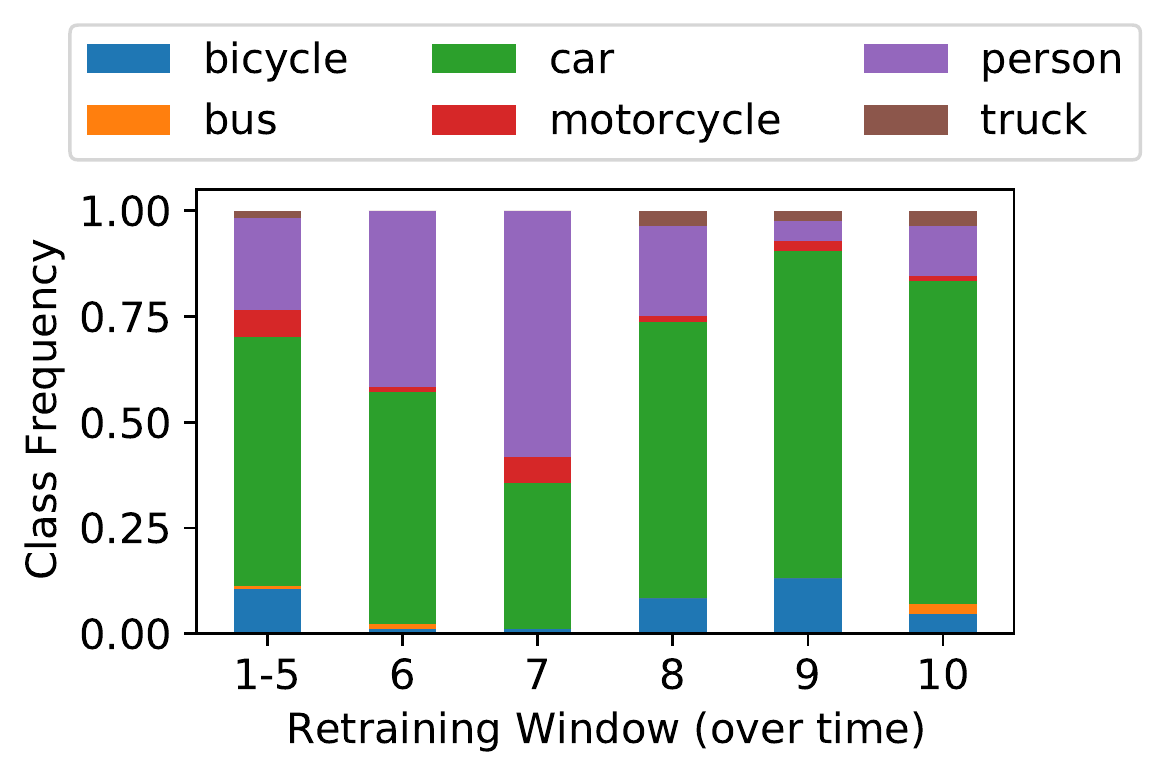}
    \caption{\small Class Distribution}
    \label{fig:jena-classdist}
  \end{subfigure}
    ~~
  \begin{subfigure}[t]{0.5\linewidth}
    \centering
    \includegraphics[width=\linewidth]{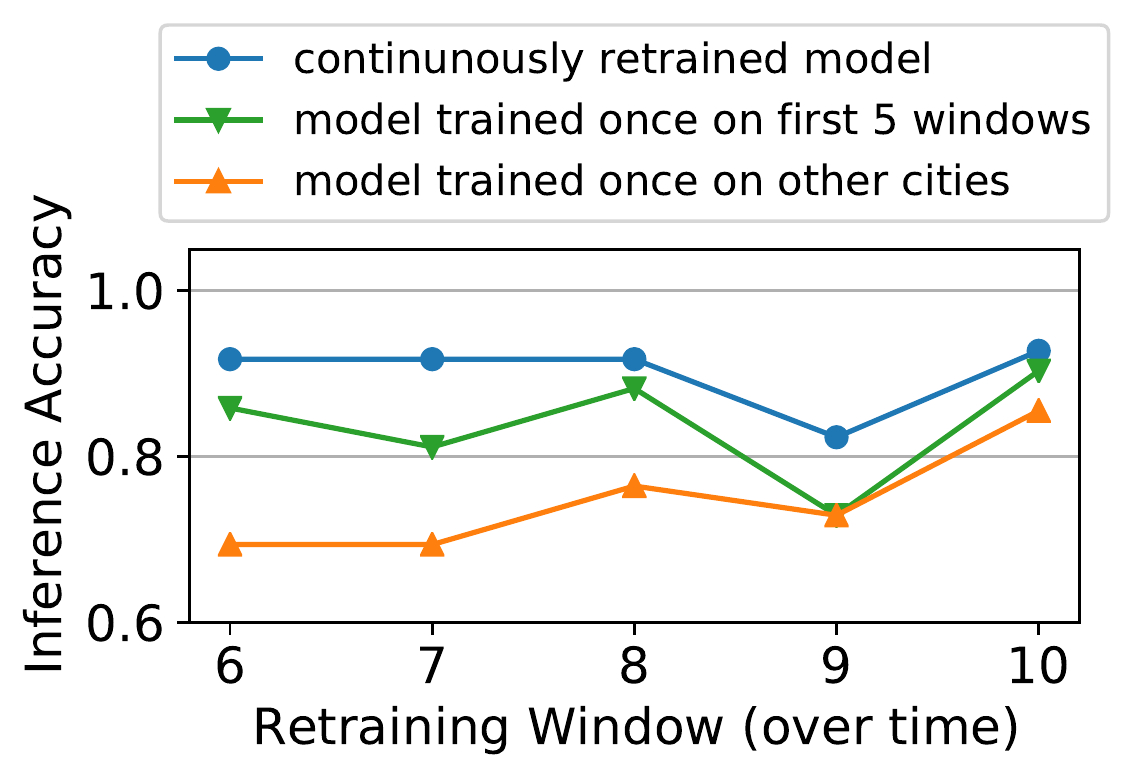}

    \caption{\small Accuracy}
    \label{fig:jena-motivation}
  \end{subfigure}
    \hspace*{\fill}
  ~~
  \begin{subfigure}[t]{0.5\linewidth}
    \centering
    \includegraphics[width=\linewidth]{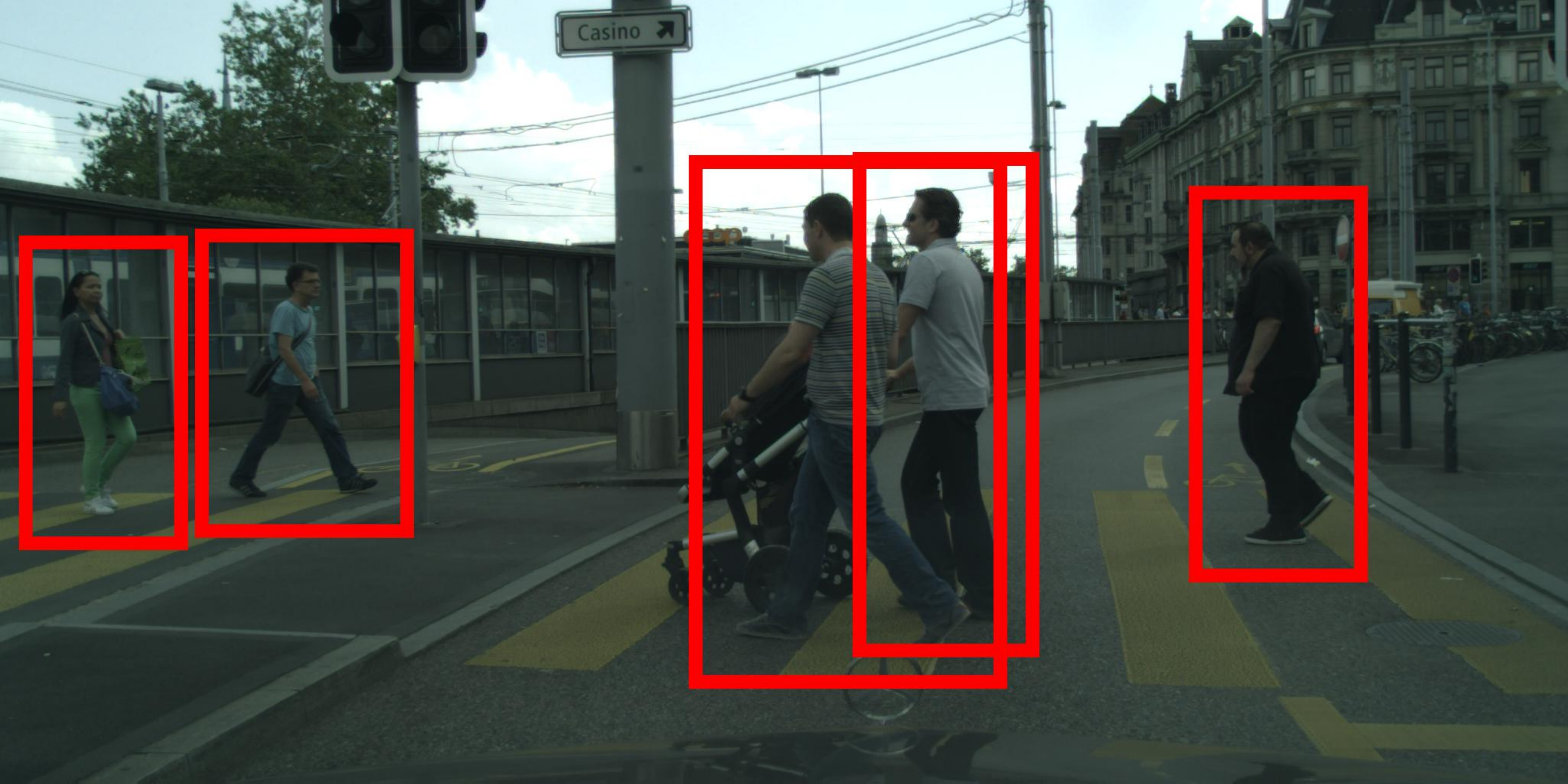}
    \caption{\small Window $1$ snapshot}
    \label{fig:jena-image-1}
  \end{subfigure}
  ~~
  \begin{subfigure}[t]{0.5\linewidth}
    \centering
    \includegraphics[width=\linewidth]{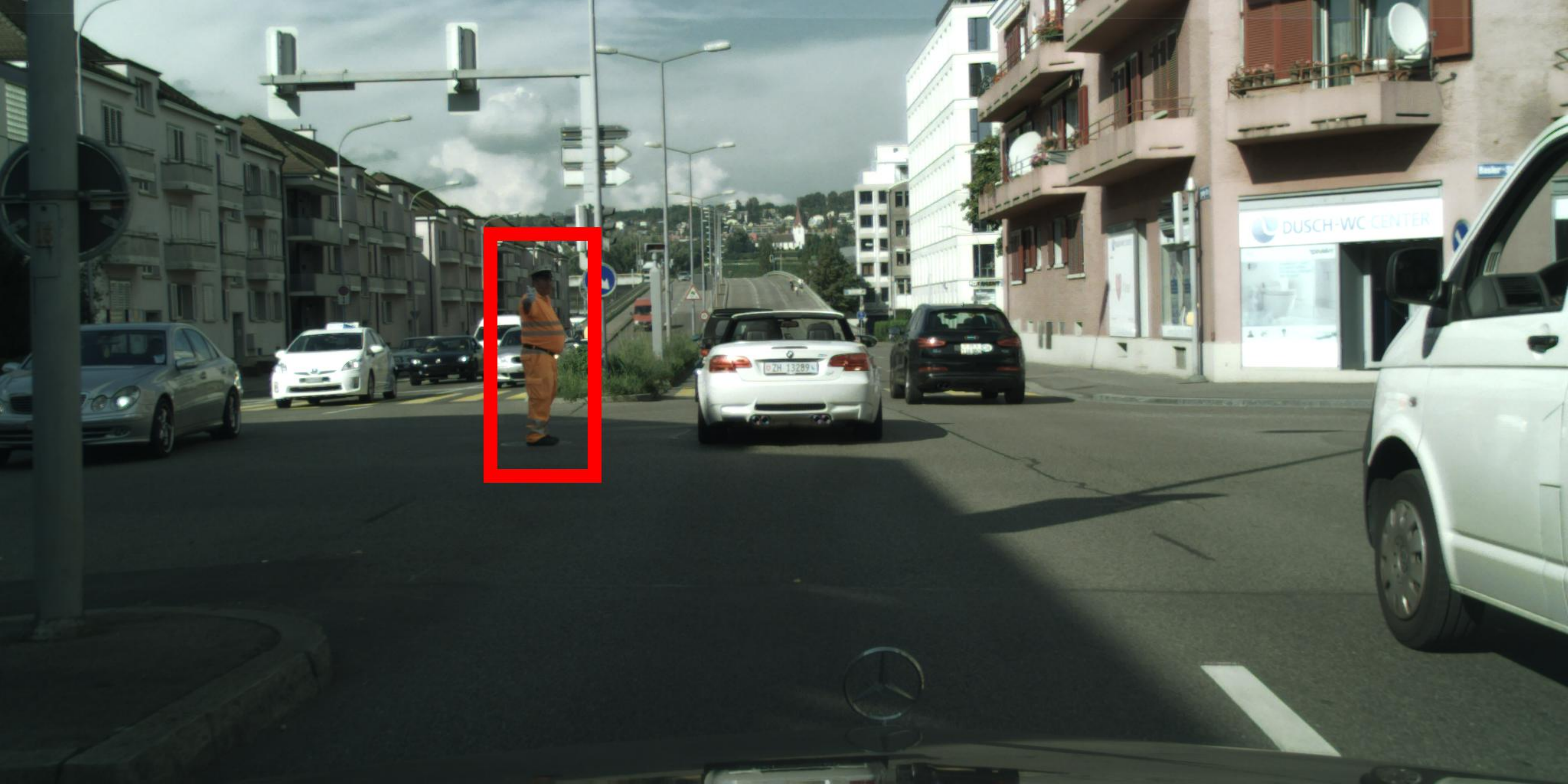}
    \caption{\small Window $9$ snapshot}
    \label{fig:jena-image-6}
  \end{subfigure}
  
   \caption{\bf\small Continuous learning in the Cityscapes data. While a shift in class distribution in window necessitates continuous learning (top-right), changes in object appearances (e.g., bottom snapshots) also influences the inference accuracy.}
  \label{fig:cityscapes-motivation}
\end{figure}


To show the benefits of continuous learning, we use the video stream from one example city in the Cityscapes dataset \cite{cityscapes} that consists of videos from dashboard cameras in many cities. 
In our evaluation in \S\ref{sec:evaluation}, we use both moving dashboard cameras as well as static cameras over long time periods. 
We divide the video data in our example city into ten fixed \emph{retraining windows} (200s in this example). 
Figure \ref{fig:jena-classdist} shows how the distribution of object classes changes among the different windows. The initial five windows see a fair amount of persons and bicycles, but bicycles rarely show up in windows 6 and 7, while the share of persons varies considerably across windows $6-10$. Even persons have different appearances (e.g., clothing and angles) over time  
(Figures \ref{fig:jena-image-1} and \ref{fig:jena-image-6}).

Figure \ref{fig:jena-motivation} plots inference accuracy of an edge DNN (a compressed ResNet18 classifier) in the last five windows using different training options. 
$(1)$ Training a compressed ResNet18 with video data on all other cities of the Cityscapes dataset does not result in good performance.
$(2)$ Unsurprisingly, we observe that training the edge DNN once using data from the first five windows {\em of this example city} improves the accuracy. 
$(3)$ 
{\em Continuous retraining} using the most recent data for training achieves the highest accuracy consistently. Its accuracy is higher than the other options 
by up to $22\%$.

Interestingly, using the data from the first five windows to train the larger ResNet101 DNN (not graphed) achieves better accuracy that nearly matches the continuously retrained ResNet18.  
The substantially better accuracy of ResNet101 compared to ResNet18 when trained {\em on the same data} of the first five windows also shows that this training data was indeed fairly representative. But the lightweight ResNet18's weights and architecture limits its ability to learn and is a key contributor to its lower accuracy.
Nonetheless, ResNet101 
is $13\times$ slower than the compressed ResNet18 \cite{cnn-perf}. 
This makes the efficient ResNet18 more suited for edge deployments and continuous learning enables it to maintain high accuracy even with data drift. Hence, the need for continuous training of edge DNNs is ongoing and not just during a ``ramp-up'' phase. 

\section{Scheduling retraining and inference jointly}
\label{sec:motivation}

We propose \emph{joint retraining and inference} on edge servers.
The joint approach utilizes resources better than statically provisioning compute for retraining on edge servers. Since retraining is periodic \cite{distribution-20, mullapudi2019} and its compute demands are far higher than inference, static provisioning causes idling and wastage.  
Compared to uploading videos to the cloud for retraining, our approach has clear advantages in privacy (\S\ref{subsec:edge}) as well as network costs and accuracy (quantified in \S\ref{subsec:eval-alternate}).

\begin{figure}[t]
  \centering
   \begin{subfigure}[t]{0.5\linewidth}
    \centering
    \includegraphics[width=\linewidth]{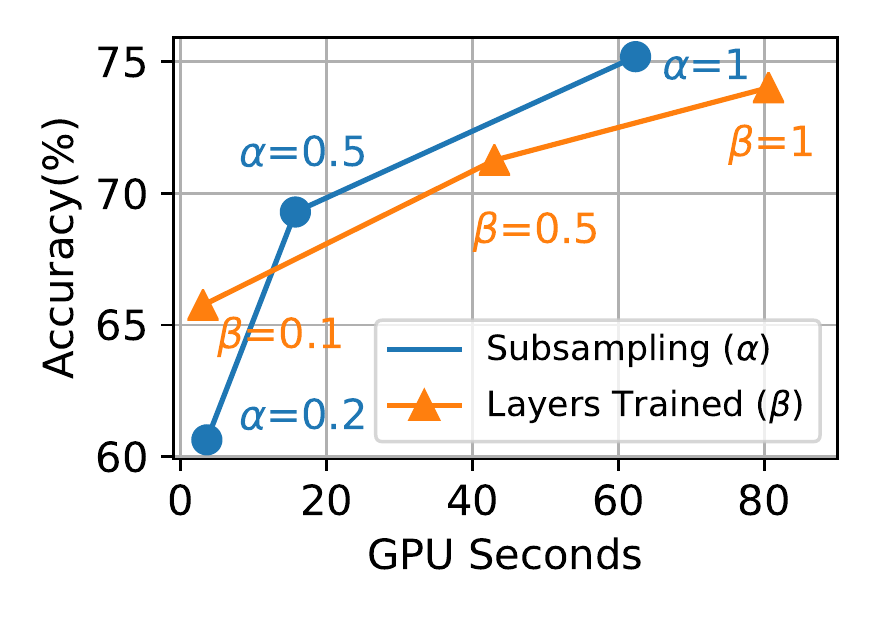}
    \caption{\small Effect of Hyperparameters}
    \label{fig:hyparam-zoom}
  \end{subfigure}
  ~~~
  \begin{subfigure}[t]{0.5\linewidth}
    \centering
    \includegraphics[width=\linewidth]{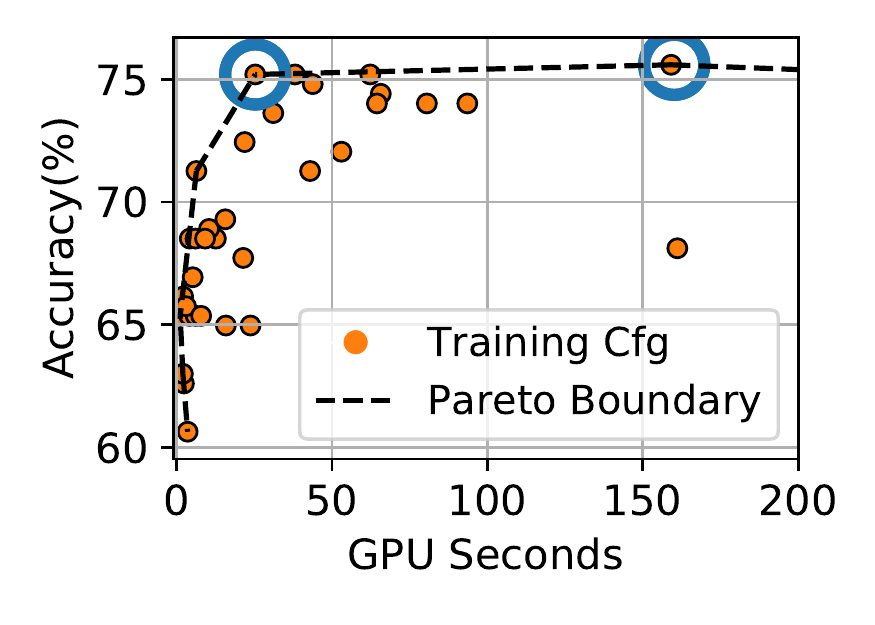}
    \caption{\small Resource-accuracy}
    \label{fig:darmstadt-profile}
  \end{subfigure}  
  \caption{\bf\small Measuring retraining configurations with the Cityscapes dataset \cite{cityscapes}. GPU seconds refers to the duration taken for retraining with $100\%$ GPU allocation. Figure \ref{fig:hyparam-zoom} varies two example hyperparameters, keeping others constant. Note the Pareto boundary of configurations in (\ref{fig:darmstadt-profile}); for every non-Pareto configuration, there is at least one Pareto configuration that is better than it in {\em both} accuracy and GPU cost. }
  \label{fig:resource-profiles}
\end{figure}

\subsection{Configuration diversity of retraining and inference}
\label{subsec:profiles}

\noindent{\bf Tradeoffs in retraining configurations.} The hyperparameters for retraining, or ``retraining configurations'', influence the resource demands and accuracy. 
Retraining fewer layers of the DNN (or, ``freezing'' more layers) consumes lesser GPU resources, as does training on fewer data samples, but they also produce a model with lower accuracy; Figure \ref{fig:hyparam-zoom}. 

Figure \ref{fig:darmstadt-profile} illustrates the resource-accuracy trade-offs for an edge DNN (ResNet18) with various hyperparameters: number of training epochs, batch sizes, number of neurons in the last layer, number of frozen layers, and fraction of training data. We make two observations.
First, there is a wide spread in the resource usage (measured in GPU seconds), by upto a factor of $200\times$. 
Second, higher resource usage does not always yield higher accuracy. For the two configurations circled in Figure \ref{fig:darmstadt-profile}, their GPU demands vary by $6\times$ even though their accuracies are the same ($\sim 76\%$). Thus, careful selection of the configurations considerably impacts the resource efficiency. However, with the characteristics of the videos changing over time, it is challenging to efficiently obtain the resource-accuracy profiles for retraining configurations.


\noindent{\bf Tradeoffs in inference configurations.} 
Inference pipelines also allow for flexibility in their resource demands at the cost of accuracy through configurations to downsize and sample frames \cite{rocket-github}. Prior work has made dramatic advancements in developing 
profilers to efficiently obtain the resource-accuracy relationship for {\em inference configurations} \cite{chameleon}. We use these efficient inference profilers in our joint solution for retraining and inference, and also to ensure that the inference pipelines continue to keep up with analyzing the live video streams with their currently allocated resources. 




\subsection{Illustrative scheduling example}
\label{subsec:motivation-sched-example}

We use an example with $3$ GPUs and two video streams, A and B, to show the considerations in scheduling inference and retraining tasks jointly.
Each retraining uses data samples accumulated since the {\em beginning of} the last retraining (referred to as the ``retraining window'').\footnote{Continuous learning targets retraining windows of durations of tens of seconds to few minutes \cite{distribution-20, mullapudi2019}. We use 120 seconds in this example. Our solution is orthogonal to the duration of the retraining window and works with any given duration in its decisions.} 
To simplify the example, we assume the scheduler has knowledge of the resource-accuracy profiles, but these are expensive to get in practice (we describe our efficient solution for profiling in \S\ref{subsec:profiling}).
Table \ref{tab:schedmot-hypparams} shows the retraining configurations ({\small Cfg1A, Cfg2A, Cgf1B, and Cgf2B}), their respective accuracies after the retraining, and GPU cost.
The scheduler is responsible for selecting configurations and allocating resources for inference and retraining jobs.

\newcommand{\scell}[2][l]{%
  \begin{tabular}[#1]{@{}c@{}}#2\end{tabular}}
  
\begin{table}[t]
\footnotesize\addtolength{\tabcolsep}{-3pt}
\begin{tabular}{l cc cc}
\toprule
\multirow{2}{*}{\bf Configuration} & \multicolumn{2}{c}{\bf Retraining Window 1} & \multicolumn{2}{c}{\bf Retraining Window 2} \\
\cmidrule(lr){2-3} \cmidrule(lr){4-5}
& \scell{\bf End\\ \bf Accuracy} & \scell{\bf GPU\\ \bf seconds} & \scell{\bf End\\ \bf Accuracy} & \scell{\bf GPU\\\bf seconds} \\ \midrule
Video A Cfg1A       & 75    & 85    & 95    & 90      \\ \midrule
Video A Cfg2A (*)   & 70    & 65    & 90    & 40     \\ \midrule
Video B Cfg1B       & 90    & 80    & 98    & 80       \\ \midrule
Video B Cfg2B (*)   & 85    & 50    & 90    & 70\\ 
\bottomrule
\end{tabular}
\caption{\label{tab:schedmot-hypparams}\small\bf Hyperparameter configurations for retraining jobs in Figure \ref{fig:schedmot}'s example. At the start of retraining window 1, camera A's inference model has an accuracy of 65\% and camera B's inference model has an accuracy of 50\%. Asterisk (*) denotes the configurations picked in Figures \ref{fig:schedmot-res-prioritization} and \ref{fig:schedmot-prioritization}.\vspace{-15pt}
}
\end{table}



\begin{figure}[t!]
  \centering
    \begin{subfigure}[t]{0.47\columnwidth}
    \centering
    \includegraphics[width=\linewidth]{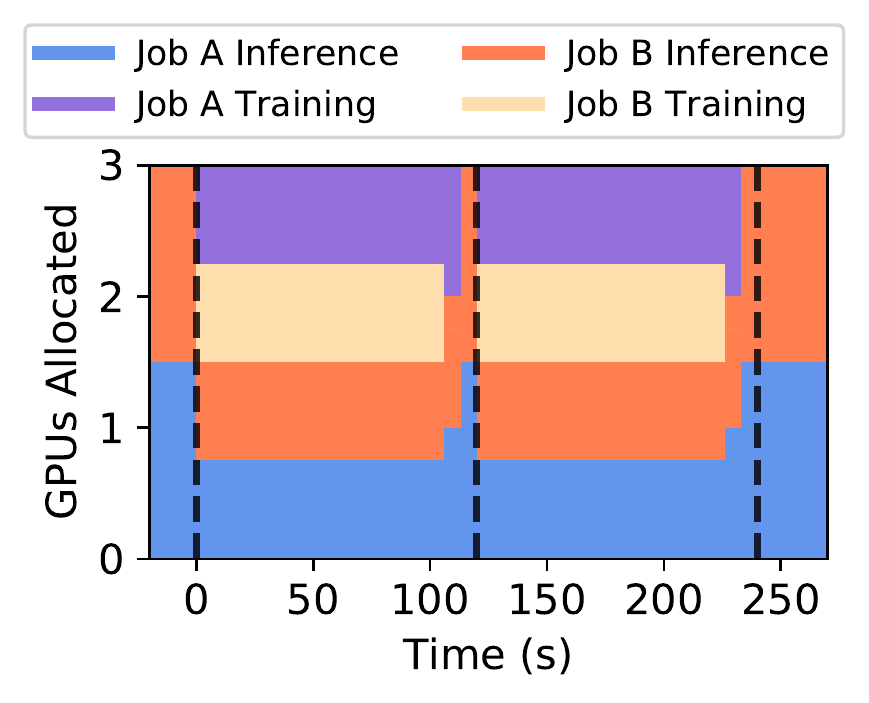}
    \caption{\small Uniform scheduler}
    \label{fig:schedmot-res-naive}
  \end{subfigure}  
  ~~
  \begin{subfigure}[t]{0.47\columnwidth}
    \centering
    \hspace*{0.05in}
    \includegraphics[width=\linewidth]{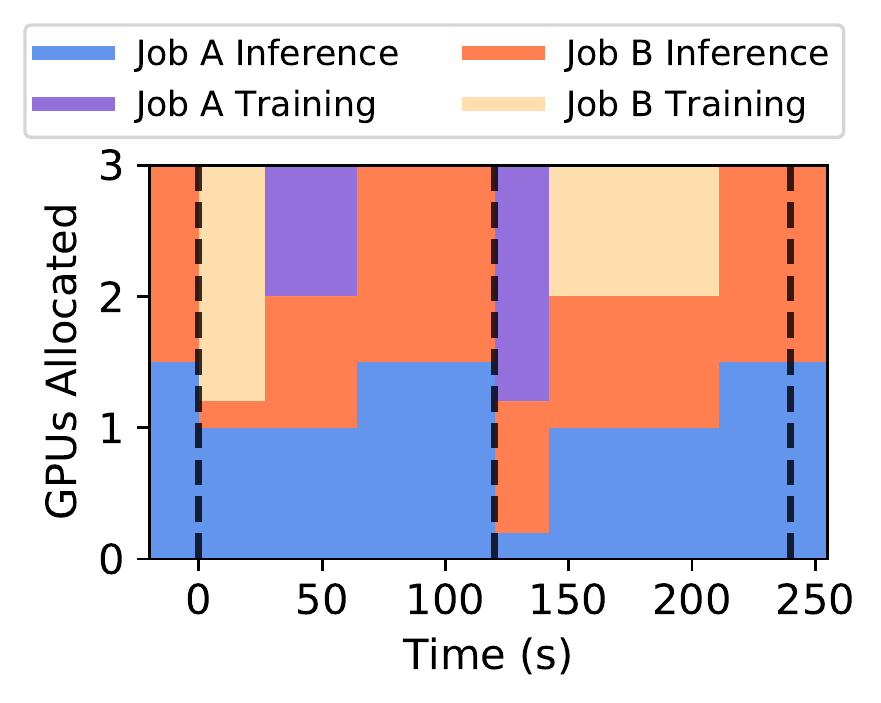}
    \caption{\small Accuracy-optimal sched.}
    \label{fig:schedmot-res-prioritization}
  \end{subfigure}
      ~~\\
  \begin{subfigure}[t]{0.5\columnwidth}
    \centering
    \includegraphics[width=\linewidth]{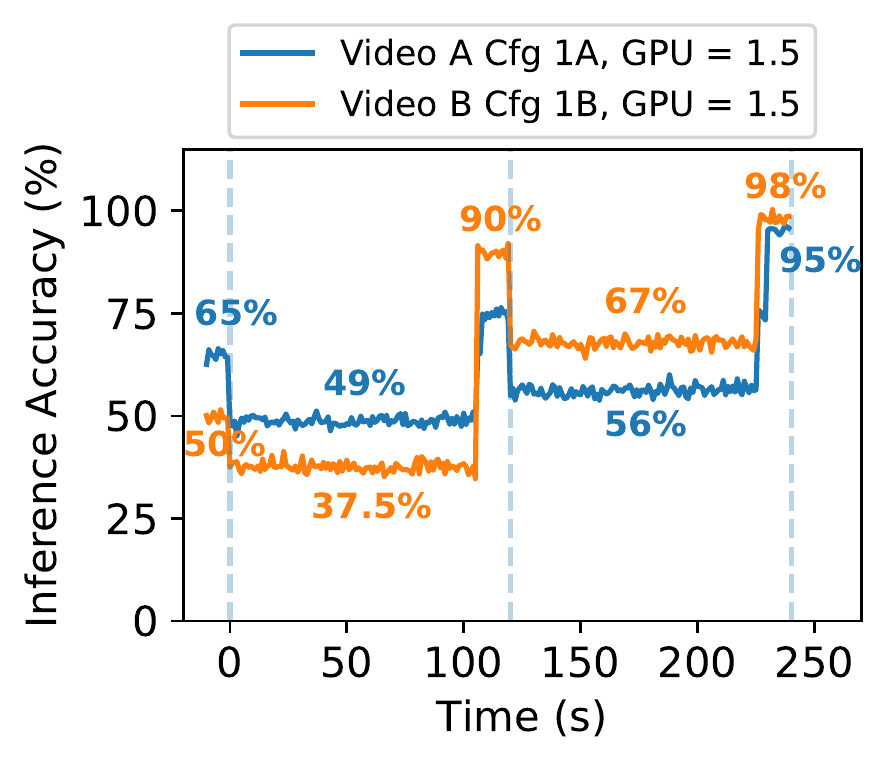}
    \caption{\small Uniform scheduler}
    \label{fig:schedmot-naive}
  \end{subfigure}  
  ~~
  \begin{subfigure}[t]{0.5\columnwidth}
    \centering
    \includegraphics[width=\linewidth]{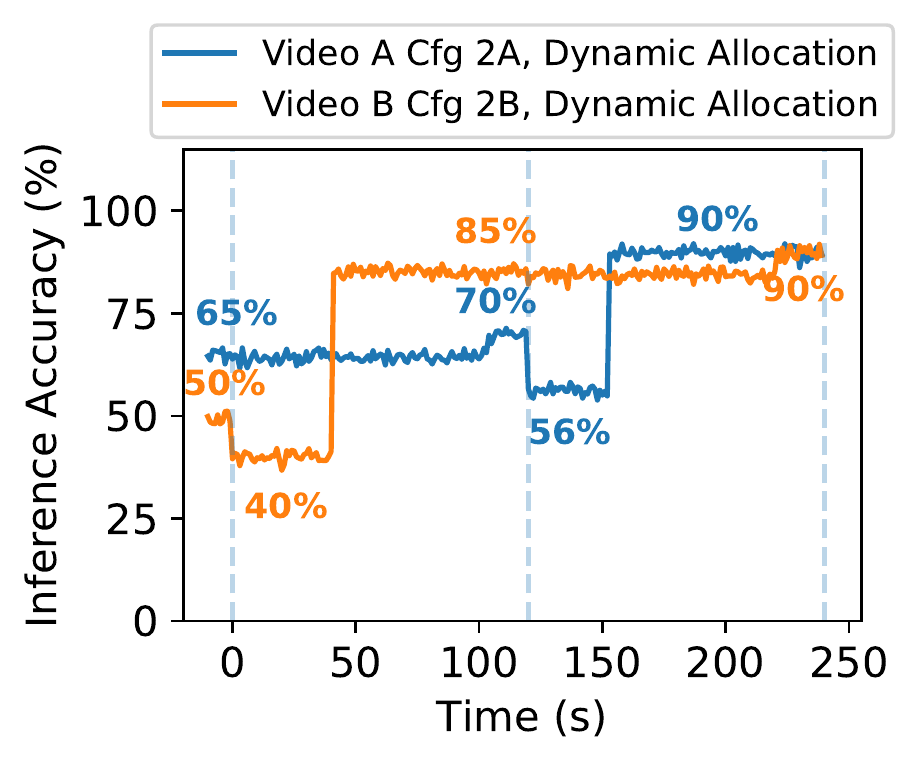}
    \caption{\small Accuracy-optimal sched.}
    \label{fig:schedmot-prioritization}
  \end{subfigure}
  ~~
  \caption{\bf\small Resource allocations (top) and inference accuracies (bottom) over time for two retraining windows (each of $120$s). The left figures show a uniform scheduler which evenly splits the $3$ GPUs, and picks configurations resulting in the most accurate models. The right figures show the accuracy-optimized scheduler that prioritizes resources and optimizes for inference accuracy averaged over the retraining window ($73\%$ compared to the uniform scheduler's $56\%$). The accuracy-optimized scheduler also ensures that inference accuracy never drops below a minimum (set to $40\%$ in this example, denoted as $a_\text{MIN}$). 
  }
  \label{fig:schedmot}
\end{figure}

\noindent{\bf {\Fair} scheduling:} Building upon prior work in cluster schedulers \cite{fair-1, fair-2} and video analytics systems \cite{videostorm}, a baseline solution for resource allocation evenly splits the GPUs between video streams, and each stream evenly partitions its allocated GPUs for retraining and inference tasks; see Figure \ref{fig:schedmot-res-naive}. Just like model training systems \cite{vizier,hyperband,pbt}, the baseline always picks the configuration for retraining that results in the highest accuracy ({\small Cfg1A, Cfg1B} for both windows).

Figure \ref{fig:schedmot-naive} shows the result of the {\fair} scheduler on the {\em inference} accuracies of the two live streams.
We see that when the retraining tasks take resources away from the inference tasks, the inference accuracy drops significantly (from {$65\%$} $\rightarrow$ {$49\%$} for video A and {$50\%$} $\rightarrow$ {$37.5\%$} for video B in Window 1).
While the inference accuracy increases significantly \emph{after} retraining, it leaves too little time in the window to reap the benefit of retraining. 
Averaged across both retraining windows, the inference accuracy across the two video streams is only $56\%$ because the gains due to the improved accuracy of the retrained model are undercut by the time taken for retraining (during which inference accuracy suffered).

\noindent{\bf Accuracy-optimized scheduling:} 
Figures \ref{fig:schedmot-res-prioritization} and \ref{fig:schedmot-prioritization} illustrate an accuracy-optimized scheduler, which by taking a holistic view on the multi-dimensional tradeoffs, provides an an average inference accuracy of $73\%$. In fact, to match the accuracies, the above {\fair} scheduler would require nearly twice the GPUs (i.e., $6$ GPUs instead of $3$ GPUs). \footnote{The techniques in our scheduler apply to other optimization metrics too, like max-min of accuracy. Evaluating other metrics is left to future work.}

This scheduler makes three key improvements. First, the scheduler selects the hyperparameter configurations based on their accuracy improvements \emph{relative} to their GPU cost. It selects lower accuracy options ({\small Cfg2A/Cfg2B}) instead of the higher accuracy ones ({\small Cfg1A/Cfg1B}) because these configurations are substantially cheaper (Table \ref{tab:schedmot-hypparams}). 
Second, the scheduler \emph{prioritizes} retraining tasks that yield higher accuracy improvement, so there is more time to reap the higher benefit from retraining. For example, the scheduler prioritizes video B's retraining in Window 1 as its inference accuracy after retraining increases by $35\%$ (compared to $5\%$ for video A).
Third, the scheduler controls the accuracy drops during retraining by balancing between the retraining time and the resources taken away from the inference tasks. 

\section{{Ekya}: Solution Description}
\label{sec:solution}





\begin{figure}[t!]
    \centering
    \includegraphics[width=.9\columnwidth]{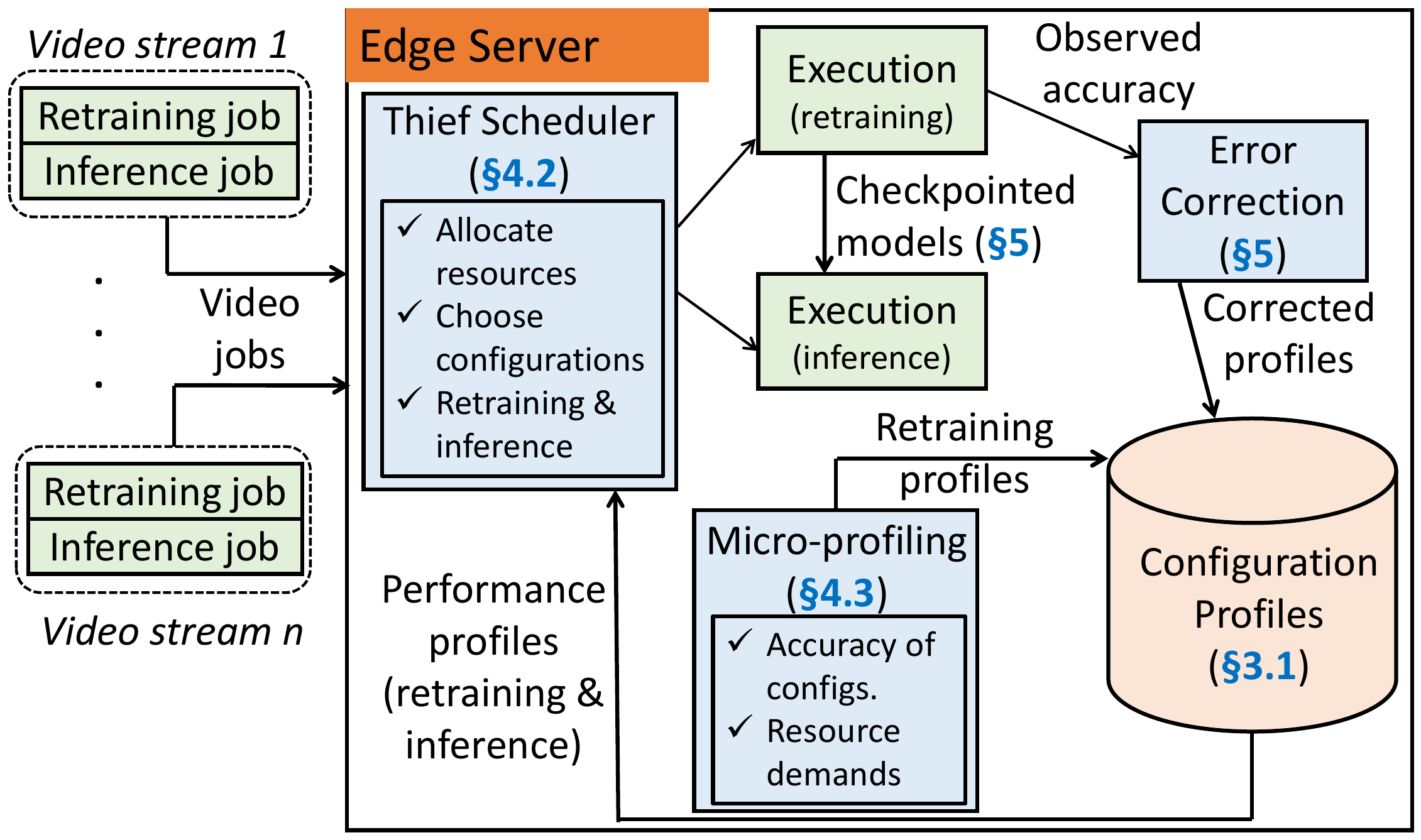}
    \caption{\small\bf {\name}'s components and their interactions. }
    \label{fig:sys-arch}
\end{figure}

Continuous training on limited edge resources requires smartly deciding when to retrain each video stream's model, how much resources to allocate, and what configurations to use. But making these decisions has two challenges.

First, the decision space of multi-dimensional configurations and resource allocations (\S\ref{subsec:profiles}) is complex. 
In fact, our problem is computationally more complex than two fundamentally challenging problems of multi-dimensional knapsack and multi-armed bandit (\S\ref{subsec:formulation}). 
Hence, we design a {\bf thief resource scheduler} (\S\ref{subsec:thief}), 
a heuristic that makes joint retraining-inference scheduling to be tractable in practice. 

Second, the decision making requires knowing the model's exact performance (in resource usage and inference accuracy), but that is difficult to obtain as it requires actually retraining using all the configurations. 
We address this challenge by designing a {\bf micro-profiling} based estimator (\S\ref{subsec:profiling}), which profiles only a few promising configurations on 
a tiny fraction of the training data with early termination.

Our techniques (in \S\ref{subsec:thief} and \S\ref{subsec:profiling}) drastically reduce the cost of decision making and performance estimation. 
Figure \ref{fig:sys-arch} presents an overview of {\name}'s components.



\subsection{Formulation of joint inference and retraining}
\label{subsec:formulation}

The problem of joint inference and retraining aims to maximize overall inference accuracy for all videos streams $\mathcal{V}$ within a given retraining window ${T}$.
(The duration of each retraining window is $\lVert T \rVert$.) 
All inference and retraining must be done in $\mathcal{G}$ GPUs.
Thus, the total compute capability is $\mathcal{G}\lVert T \rVert$ GPU-time. Without loss of generality, let $\delta$ be the smallest granularity of GPU allocation.  
Table \ref{tab:notations} lists the notations. 
As explained in \S\ref{subsec:profiles}, each video $v \in V$ has a set of \emph{retraining} configurations $\Gamma$ 
and a set of \emph{inference} configurations $\Lambda$. 

\noindent\textbf{Decisions.} For each video $v\in\mathcal{V}$ in a window $T$, we decide: (1) the retraining configuration $\gamma\in\Gamma$ ($\gamma = \emptyset$ means no retraining); (2) the inference configuration $\lambda\in\Lambda$; and (3) how many GPUs (in multiples of $\delta$) 
to allocate for retraining ($\mathcal{R}$) 
and inference ($\mathcal{I}$). 
We use binary variables $\phi_{v\gamma\lambda\mathcal{R}\mathcal{I}}\in\{0,1\}$ to denote these decisions (see Table \ref{tab:notations} for the definition). 
These decisions require $C_T(v, \gamma, \lambda)$ GPU-time and yields overall accuracy of $A_T(v, \gamma, \lambda, \mathcal{R}, \mathcal{I})$. As we saw in \S\ref{subsec:motivation-sched-example}, $A_T(v, \gamma, \lambda, \mathcal{R}, \mathcal{I})$ is averaged across the window $T$, and the inference accuracy at {\em each point in time} is determined by the above decisions.



\begin{table}[t!]
\footnotesize
\begin{tabular}{cl}
{\bf Notation} & {\bf Description}\\\hline
$\mathcal{V}$ & Set of video streams\\
$v$ & A video stream ($v \in \mathcal{V}$)\\\hline
$T$ & A retraining window with duration $\lVert T \rVert$ \\\hline
$\Gamma$ & Set of all retraining configurations\\
$\gamma$ & A retraining configuration ($\gamma \in \Gamma$)\\\hline
$\Lambda$ & Set of all inference configurations\\
$\lambda$ & An inference configuration ($\lambda \in \Lambda$)\\\hline
$\mathcal{G}$ & Total number of GPUs\\
$\delta$ & The unit for GPU resource allocation \\\hline
$A_T(v, \gamma, \lambda, \mathcal{R}, \mathcal{I} )$ & Inference accuracy for video $v$ for\\
                                   &given configurations and allocations\\
$C_T(v, \gamma, \lambda)$ & Compute cost in GPU-time for video $v$ for\\
                                   &given configurations and allocations\\\hline
$\phi_{v\gamma\lambda\mathcal{R}\mathcal{I}}$ & A set of binary variables ($\phi_{v\gamma\lambda\mathcal{R}\mathcal{I}}\in\{0,1\}$). \\
& $\phi_{v\gamma\lambda\mathcal{R}\mathcal{I}} = 1$ iff we use retraining config $\gamma$, \\
&inference config $\lambda$, $\mathcal{R}\delta$ GPUs for retraining,\\
& $\mathcal{I}\delta$ GPUs for inference for video $v$\\\hline
\end{tabular}
\caption{\label{tab:notations}\small\bf Notations used in {\name}'s description.}\vspace{-12pt}
\end{table}

\noindent\textbf{Optimization Problem.} The optimization problem maximizes inference accuracy averaged across all videos in a retraining window within the GPU resource limit. 

{\small
\vspace{-12pt}
\begin{equation}
    \begin{aligned}
       & \underset{\phi_{v\gamma\lambda\mathcal{R}\mathcal{I}}}{\arg\max}
         \frac{1}{\lVert \mathcal{V} \rVert}
         \sum_{\substack{\forall v\in\mathcal{V}, 
                        \forall \gamma\in\Gamma,
                        \forall \lambda\in\Lambda,\\
                        \forall \mathcal{R}, \forall \mathcal{I} \in \{0, 1, ..., \frac{\mathcal{G}}{\delta}\} 
                        }}
          \phi_{v\gamma\lambda\mathcal{R}\mathcal{I}} \cdot
          A_T(v, \gamma, \lambda, \mathcal{R}, \mathcal{I})\\
       & \text{subject to}\\
       & 1. \sum_{\substack{\forall v\in\mathcal{V}, 
                        \forall \gamma\in\Gamma,
                        \forall \lambda\in\Lambda,\\
                        \forall \mathcal{R}, \forall \mathcal{I}}}
                        \phi_{v\gamma\lambda\mathcal{R}\mathcal{I}} \cdot
                        C_T(v, \gamma, \lambda)
                        \leq \mathcal{G}\lVert T \rVert \\
      & 2. \sum_{\substack{\forall v\in\mathcal{V},
                        \forall \gamma\in\Gamma,
                        \forall \lambda\in\Lambda,\\
                        \forall \mathcal{R}, \forall \mathcal{I}}}
                           \phi_{v\gamma\lambda\mathcal{R}\mathcal{I}} \cdot
                           (\mathcal{R} + \mathcal{I})
                        \leq \frac{\mathcal{G}}{\delta} \\
      & 3. \sum_{\substack{\forall \gamma\in\Gamma, 
                           \forall \lambda\in\Lambda,\\ 
                           \forall \mathcal{R}, 
                           \forall \mathcal{I}}} 
           \phi_{v\gamma\lambda\mathcal{R}\mathcal{I}} \leq 1, 
           \forall v\in\mathcal{V}
    \end{aligned}
    \label{eqn:optimization}
\end{equation}
}%

The first constraint ensures that the GPU allocation does not exceed the available GPU-time $\mathcal{G}\lVert T \rVert$ available in the retraining window. The second constraint ensures that {\em at any point in time} the GPU allocation (in multiples of $\delta$) does not exceed the total number of available GPUs. 
The third constraint ensures that at most one retraining configuration and one inference configuration are picked for each video $v$.

\noindent\textbf{Complexity Analysis.} \emph{Assuming} all the $A_T(v, \gamma, \lambda, \mathcal{R}, \mathcal{I})$ values are known, the above optimization problem can be reduced to a multi-dimensional binary knapsack problem, a NP-hard problem \cite{DBLP:journals/mor/MagazineC84}. 
Specifically, the optimization problem is to pick binary options ($\phi_{v\gamma\lambda\mathcal{R}\mathcal{I}}$) to maximize overall accuracy while satisfying two capacity constraints (the first and second constraints in Eq~\ref{eqn:optimization}). 
In practice, however, getting all the $A_T(v, \gamma, \lambda, \mathcal{R}, \mathcal{I})$ is \emph{infeasible} 
because this requires training the edge DNN using all retraining configurations and running inference using all the retrained DNNs with all possible GPU allocations and inference configurations.

The uncertainty of $A_T(v, \gamma, \lambda, \mathcal{R}, \mathcal{I})$ resembles the multi-armed bandits (MAB) problem \cite{robbins1952some} to maximize the expected rewards given a limited number of trials for a set of options.
Our optimization problem is more challenging than MAB for two reasons.
First, unlike the MAB problem, the cost of trials ($C_T(v, \gamma, \lambda)$) varies significantly, and the optimal solution may need to choose cheaper yet less rewarding options to maximize the overall accuracy.
Second, getting the reward $A_T(v, \gamma, \lambda, \mathcal{R}, \mathcal{I})$ after each trial 
requires "ground truth" labels that are obtained using the large golden model, which can only be used judiciously on resource-scarce edges (\S\ref{subsec:continuous}).

In summary, our optimization problem is computationally more complex than two fundamentally challenging problems (multi-dimensional knapsack and multi-armed bandits).  




\subsection{Thief Scheduler}
\label{subsec:thief}

Our scheduling heuristic makes the scheduling problem tractable by decoupling resource allocation (i.e., $\mathcal{R}$ and $\mathcal{I}$) and configuration selection (i.e., $\gamma$ and $\lambda$) (Algorithm \ref{algo:thief_sched}). 
We refer to {\name}'s scheduler as the ``thief'' scheduler and it iterates among all inference and retraining jobs as follows.

{\bf (1)} It starts with a fair allocation for all video streams $v \in V$ (line 2 in Algorithm \ref{algo:thief_sched}). 
In each step, it iterates over all the inference and retraining jobs of each video stream (lines 5-6), and {\em steals} a tiny quantum $\Delta$ of resources (in multiples of $\delta$; see Table \ref{tab:notations}) from each of the other jobs (lines 10-11).

{\bf (2)} With the new resource allocations ({\small temp\_alloc[]}), it then selects configurations for the jobs using the {\sf\footnotesize PickConfigs} method (line 14 and Algorithm \ref{algo:pickconfigs}) that iterates over all the configurations for inference and retraining for each video stream.  
For inference jobs, among all the configurations whose accuracy is $\geq a_\text{MIN}$, {\sf\footnotesize PickConfigs} picks the configuration with the highest accuracy that can keep up with the inference of the live video stream within the current allocation (line 3-4 in Algorithm \ref{algo:pickconfigs}).  

For retraining jobs, {\sf\footnotesize PickConfigs} picks the configuration that maximizes the accuracy $A_T(v, \gamma, \lambda, \mathcal{R}, \mathcal{I})$ over the retraining window for each video $v$ (lines 6-12 in Algorithm \ref{algo:pickconfigs}). {\sf\footnotesize EstimateAccuracy} in Algorithm \ref{algo:pickconfigs} (line 7) aggregates the instantaneous accuracies over the retraining window for a given pair of inference configuration (chosen above) and retraining configuration. {\name}'s micro-profiler (\S\ref{subsec:profiling}) provides the estimate of the accuracy and the time to retrain for a retraining configuration when $100\%$ of GPU is allocated, and {\sf\footnotesize EstimateAccuracy} proportionately scales the GPU-time for the current allocation (in {\small temp\_alloc[]}) and training data size. In doing so, it avoids configurations whose retraining durations exceed $\lVert T \rVert$ with the current allocation (first constraint in Eq. \ref{eqn:optimization}). 


{\bf (3)} After reassigning the configurations, {\name} uses the estimated average inference accuracy (accuracy\_avg) over the retraining window (line 14 in Algorithm \ref{algo:thief_sched}) and keeps the new allocations only if it improves up on the accuracy from prior to stealing the resources (line 15 in Algorithm \ref{algo:thief_sched}).


The thief scheduler repeats the process till the accuracy stops increasing (lines 15-20 in Algorithm \ref{algo:thief_sched}) and until all the jobs have played the ``thief''. 
Algorithm~\ref{algo:thief_sched} is invoked at the beginning of each retraining window, as well as on the completion of every training job during the window to reallocate resources to the other training and inference jobs.

%
%

\begin{algorithm}[t]
\small
 \KwData{Training ($\Gamma$) and inference ($\Lambda$) configurations}
 \KwResult{GPU allocations $\mathcal{R}$ and $\mathcal{I}$, chosen configurations ($\gamma \in \Gamma$, $\lambda \in \Lambda$) $\forall v \in V$}
 
all\_jobs[] = Union of inference and training jobs of videos $V$\;
\tcc{Initialize with fair allocation}
 best\_alloc[] = fair\_allocation(all\_jobs)\; 
 best\_configs[], best\_accuracy\_avg =  {\sf\footnotesize PickConfigs}(best\_alloc)\;
     \tcc{Thief resource stealing}
     \For{\text{\em thief\_job} in \text{\em all\_jobs[]}}{
        \For{\text{\em victim\_job} in \text{\em all\_jobs[]}}{
            \If{\text{\em thief\_job} == \text{\em victim\_job}} {\bf continue\;}
            temp\_alloc[] $\leftarrow$ best\_alloc[]\;
            \While{true}{
                \tcc{$\Delta$ is the increment of stealing}
                temp\_alloc[victim\_job] $-$= $\Delta$\;
                temp\_alloc[thief\_job] $+$= $\Delta$\;
                \If{{\em temp\_alloc[victim\_job]} < {\em 0}}{
                    \bf break\;
                }
                \tcc{Calculate accuracy over retraining window and pick configurations.}temp\_configs[], accuracy\_avg = {\sf\footnotesize PickConfigs}(temp\_alloc[])\;
                \If{{\em accuracy\_avg} > {\em best\_accuracy\_avg}}{
                    best\_alloc[] = temp\_alloc[]\;
                    best\_accuracy\_avg = accuracy\_avg\;
                    best\_configs[] = temp\_configs[];
                }
                \Else{\bf break\;}
            }
        }
     }
 {\bf return} {best\_alloc[], best\_configs[]}\;
 
 \caption{\bf\small Thief Scheduler.
 }
 \label{algo:thief_sched}
\end{algorithm}


\begin{algorithm}[t]
\small
 \KwData{Resource allocations in {temp\_alloc[]}, configurations ($\Gamma$ and $\Lambda$), retraining window $T$, videos $V$}
 \KwResult{Chosen configs $\forall v \in V$, 
 average accuracy over $T$}
 
     chosen\_accuracies[] $\leftarrow$\{\}; chosen\_configs[] $\leftarrow$\{\}\;
     \For{\text{\em v} in $V$\text{\em []}}{
        infer\_config\_pool[] = $\Lambda$.{\bf where}(\text{resource\_cost} < temp\_alloc[v.inference\_job] \&\& accuracy $\geq a_\text{MIN}$ )\;
        infer\_config = {\bf max}(infer\_config\_pool, {\bf key}=accuracy)\;
        best\_accuracy = 0\;
        \For{\text{\em train\_config} in \text{\em $\Gamma$}}{
            \tcc{Estimate accuracy of inference/training config pair over retraining window} 
            accuracy = {\sf\footnotesize EstimateAccuracy}(train\_config, infer\_config, temp\_alloc[v.training\_job], $T$)\;
            \If{\text{\em accuracy} > \text{\em best\_accuracy}} {
                best\_accuracy = accuracy\;
                best\_train\_config = train\_config\;
            }
        }
        chosen\_accuracies[v] = best\_accuracy\;
        chosen\_configs[v] = \{infer\_config, best\_train\_config\}\;
     }
    {\bf return} chosen\_configs[], {\bf mean}(chosen\_accuracies[])\;
 
 \caption{\bf\small PickConfigs}
 \label{algo:pickconfigs}
\end{algorithm}



\mypara{Design rationale} 
We call out the key aspects that makes the scheduler's decision efficient by pruning the search space. 
\begin{packeditemize}

\item {\em Coarse allocations:}
The thief scheduler allocates GPU resources in quantums of $\Delta$. We empirically pick the value of $\Delta$ that is coarse and yet accurate enough for scheduling decisions,  while also being mindful of the granularity achievable in modern GPUs \cite{nvidia-mps}. We analyze the sensitivity of $\Delta$ in \S\ref{subsec:eval-understanding}. Resource stealing always ensures that the total allocation is within the limit (second constraint in Eq~\ref{eqn:optimization}). 


\item {\em Reallocating resources only when a retraining job completes:}
Although one can reallocate GPU resource among jobs at finer temporal granularity (\eg whenever a retraining job has reached a high accuracy), we empirically find that the gains from such complexity is marginal. 
That said, \name periodically checkpoints the model (\S\ref{sec:system}) so that inference can get the up-to-date accuracy from retraining.

\item{\em Pruned configuration list:} 
Our micro-profiler (described next) speeds up the thief scheduler by giving it only the more promising configurations. Thus, the list $\Gamma$ used in Algorithm \ref{algo:thief_sched} is significantly smaller than the exhaustive set.

\end{packeditemize}


\subsection{Performance estimation with micro-profiling}
\label{subsec:profiling}

{\name}'s scheduling decisions in \S\ref{subsec:thief} rely on estimations of post-retraining accuracy and resource demand of the retraining configurations. 
Specifically, at the beginning of each retraining window $T$, we need to {\em profile} for each video $v$ and each configuration $\gamma\in\Gamma$, the accuracy after retraining using $\gamma$ and the corresponding time taken to retrain.

\mypara{Profiling in \name vs. hyperparameter tuning}
While \name's profiling may look similar to hyperparameter tuning (e.g.,~\cite{DBLP:conf/nips/SnoekLA12,DBLP:journals/jmlr/LiJDRT17}) at first blush, there are two key differences. 
First, \name needs the performance estimates of a broad set of candidate configurations for the thief scheduler, not just of the single best configuration, because the best configuration is jointly decided across the many retraining and inference jobs. 
Second, in contrast to hyperparameter tuning which runs separately of the eventual inference/training, \name's profiling must share compute resource with all retraining and inference.

One strawman is to predict the performance of configurations based on their history from prior training instances, but we have found that this works poorly in practice. 
In fact, even when we cached and reused models from prior retraining windows with {\em similar} class distributions, the accuracy was still substantially lower due to other factors that are difficult to model like lighting, angle of objects, density of the scene, etc. (see \S\ref{subsec:eval-alternate}). Thus we adopt an {\em online} approach for estimation by using the current retraining window's data.


\mypara{Opportunities} 
\name leverages three empirical observations for efficient profiling of the retraining configurations. 
$(i)$ Resource demands of the configurations are deterministic. Hence, we measure the GPU-time taken to retrain for {\em each epoch} in the current retraining window when $100\%$ of the GPU is allocated to the retraining. This allows us to scale the time for varying number of epochs, GPU allocations, and training data sizes in Algorithm \ref{algo:thief_sched}. 
$(ii)$ Post-retraining accuracy can be roughly estimated by training on a small subset of training data for a handful of epochs.
$(iii)$ The thief scheduler's decisions are not impacted by small errors in the estimations.



\mypara{Micro-profiling design} 
The above insights inspired our approach, called {\em micro-profiling}, where 
for each video, we test various retraining configurations, but on a {\em small subset} of the retraining data and only for a {\em small number} of epochs (well before models converge). 
Our micro-profiler is nearly $100\times$ more efficient than exhaustive profiling (of all configurations on the entire training data), while predicting accuracies with an error of $5.8\%$, 
which is low enough in practice to {\em mostly} ensure that the thief scheduler makes the same decisions as it would with a fully accurate prediction. 
We explain the techniques that make {\name}'s micro-profiling efficient.


\noindent{\em 1) Training data sampling:}  
{\name}'s micro-profiling works on only a small fraction (say, $5\%-10\%$) of the training data in the retraining window (which is already a subset of all the videos accumulated in the retraining window). While we considered weighted sampling techniques for the micro-profiling, we find that uniform random sampling is the most indicative of the configuration's performance on the full training data, since it preserves all the data distributions and variations. 

\noindent{\em 2) Early termination:} 
Similar to data sampling, {\name}'s micro-profiling only tests each configuration for a small number (say, 5) of training epochs.
Compared to a full fledged profiling that needs few tens of epochs to converge, such early termination greatly speeds up the micro-profiling process.

After early termination on the sampled training data, we obtain the (validation) accuracy of each configuration at each epoch it was trained. We then fit the accuracy-epoch points to the a non-linear curve model from \cite{optimus} using a non-negative least squares solver~\cite{nnls}. This model is then used to extrapolate the accuracy that would be obtained by retraining with all the data for larger number of epochs. The use of this extrapolation is consistent with similar work in this space~\cite{themis,optimus}. 

\noindent{\em 3) Pruning out bad configurations:} 
Finally, {\name}'s micro-profiling also prunes out those configurations for micro-profiling (and hence, for retraining) that have historically not been useful. These are configurations that are usually significantly distant from the configurations on the Pareto curve of the resource-accuracy profile (see Figure \ref{fig:darmstadt-profile}), and thus unlikely to be picked by the thief scheduler. Avoiding these configurations improves the efficiency of the micro-profiling. 

\mypara{Annotating training data} 
For both the micro-profiling as well as the retraining, {\name} acquires labels using a ``golden model'' (\S\ref{subsec:continuous}). This is a high-cost but high-accuracy model trained on a large dataset. 
As explained in \S\ref{sec:background}, the golden model cannot keep up with inference on the live videos and we use it to label only a small subset of the videos for retraining.

\section{Ekya Implementation}
\label{sec:system}







\name uses PyTorch \cite{pytorch} for running and training ML models. 

\mypara{Modularization}
Our implementation uses a collection of logically distributed modules for ease of scale-out to many video streams and resources. 
Each module acts as either the \name scheduler, micro-profiler, or a training/inference job, and is implemented by a long-running ``actor'' in Ray \cite{ray}. 
A benefit of using the actor abstraction is its highly optimized initialization cost and failure recovery. 

\mypara{Dynamic reallocation of resources}
\name reallocates GPU resources between training and inference jobs at timescales that are far more dynamic than required by prior frameworks (where the GPU allocations for jobs are fixed upfront ~\cite{kubernetes, yarn}).  
While a middle layer like Nvidia MPS~\cite{nvidia-mps} provides resource isolation in the GPU by intercepting CUDA calls and re-scheduling them, it also 
requires terminating and restarting a process to change its resource allocation. 
Despite this drawback, \name uses Nvidia MPS due to its practicality, while the restarting costs are largely avoided by the actor-based implementation that keeps DNN model in GPU memory. 

\mypara{Placement onto GPUs} 
The resource allocations produced by the thief scheduler are ``continuous'', i.e., it assumes that the fractional resources can be spanned across two discrete GPUs. To avoid the consequent expensive inter-GPU communication, \name first quantizes the allocations to inverse powers of two (\eg 1/2, 1/4, 1/8). This makes the jobs amenable to packing. \name then allocates jobs to GPUs in descending order of demands to reduce fragmentation \cite{tetris}. 


\mypara{Model checkpointing and reloading}
\name can improve inference accuracy by checkpointing the model {\em during} retraining and dynamically loading it as the inference model~\cite{tf-checkpoint, torch-checkpoint}.
Checkpointing can, however, disrupt both the retraining and the inference jobs, 
so {\name} weighs the cost of the disruption (\ie additional delay on retraining and inference) due to checkpointing against its benefits (\ie the more accurate model is available sooner). 
Implementing checkpointing in \name is also made easy by the actor-based programming model that allows for queuing of requests when the actor (model) is unavailable when its new weights are being loaded. 

\mypara{Adapting estimates during retraining}
When the accuracy during the retraining varies from the expected value from micro-profiling, {\name} reactively adjusts its allocations. 
Every few epochs, \name uses the current accuracy of the model being retrained to estimate its eventual accuracy when all the epochs are complete. It updates the expected accuracy in the profile of the retraining ($\Gamma$) with the new value, and then reruns Algorithm \ref{algo:thief_sched} for new resource allocations (but leaves the configuration that is used currently, $\gamma$, to be unchanged). 
\section{Evaluation}
\label{sec:evaluation}

We evaluate \name's performance and the key findings are:

\noindent{\bf 1)} Compared to static retraining baselines, \name achieves upto 29\% higher accuracy. For the baseline to match \name's accuracy, it would require $4\times$ additional GPU resources. (\S\ref{subsec:eval:overall}) 

\noindent{\bf 2)} Both micro-profiling and thief scheduler contribute sizably to \name's gains. (\S\ref{subsec:eval-understanding}) 
In particular, the micro-profiler estimates accuracy with low median errors of $5.8\%$. 
(\S\ref{subsec:eval-profiling})

\noindent{\bf 3)} The thief scheduler efficiently makes its decisions in 9.4s when deciding for 10 video streams across 8 GPUs with 18 configurations per model for a 200s retraining window. (\S\ref{subsec:eval-understanding})

\noindent{\bf 4)} Compared to alternate designs, including retraining the models in the cloud or using pre-trained cached models, \name achieves a higher accuracy without the network costs. (\S\ref{subsec:eval-alternate})

\subsection{Setup}
\label{subsec:eval-setup}


\mypara{Datasets} 
We use both on-road videos captured by dashboard cameras as well as urban videos captured by mounted cameras. The dashboard camera videos are from cars driving through cities in the US and Europe, Waymo Open~\cite{waymo} (1000 video segments with in total 200K frames) and Cityscapes~\cite{cityscapes} (5K frames captured by 27 cameras) videos. The urban videos are from stationary cameras mounted in a building (``Urban Building'') as well as from five traffic intersections (``Urban Traffic''), both collected over 24-hour durations. 
We use a retraining window of 200 seconds in our experiments, and split each of the videos into 200 second segments.  
Since the Waymo and Cityscapes dataset do not contain continuous timestamps, we 
create retraining windows by concatenating images from the same camera in chronological order to form a long video stream and split it into 200 second segments. 

\mypara{DNN models.} We use the ResNet18 object classifier model as our edge DNN. As explained in \S\ref{subsec:continuous}, we use an expensive golden model (ResNeXt 101 
\cite{wang2019elastic}) to get ground truth labels for training and testing. 
On a subset of data that have human annotations, we confirm that the labels produced by the golden model are very similar to human-annotated labels. 





\mypara{Testbed and trace-driven simulator}
We run \name's implementation (\S\ref{sec:system}) on AWS EC2 p3.2xlarge instances for 1 GPU experiments and  p3.8xlarge instances for 2 GPU experiments. Each instance has Nvidia V100 GPUs with NVLink interconnects and Intel Skylake Xeon processors.   

We also built a simulator to 
test \name under a wide range of resource constraints, workloads, and longer durations. 
The simulator takes as input the accuracy and resource usage (in GPU time) of training/inference configurations logged from our testbed. 
For each training job in a window, we log the training-accuracy progression over GPU-time. We also log the inference accuracy on the real videos to replay it in our simulator. 
This exhaustive trace allows us to mimic the jobs with high fidelity under different scheduling policies. 


\mypara{Retraining configurations}
As listed in \S\ref{subsec:profiles}, our retraining configurations are obtained by combining these hyperparameters: number of epochs to train, batch size, number of neurons in the last layer, number of layers to retrain, and the fraction of data between retraining windows to use for retraining.



\mypara{Baselines}
Our baseline, called {\em \fair scheduler},  
uses $(a)$ a fixed retraining configuration, and $(b)$ a static retraining/inference resource allocation (these are adopted by prior schedulers \cite{fair-1, fair-2, videostorm}). 
For each dataset, we test all retraining configurations on a hold-out dataset \footnote{
The same hold-out dataset is used to customize the off-the-shelf DNN inference model. This is a common strategy in prior work (\eg~\cite{noscope}).} (\ie two video streams that were never used in later tests) to produce the Pareto frontier of the accuracy-resource tradeoffs 
(\eg Figure~\ref{fig:resource-profiles}). 
The \fair scheduler then picks two points on the Pareto frontier as the fixed retraining configurations to represent ``high'' (Config 1)  and ``low'' (Config 2) resource usage, and uses one of them for all retraining windows in a test.

We also consider two alternative designs. 
(1) {\em offloading retraining to the cloud}, and 
(2) {\em caching and re-using a retrained model} from history. 
We will present their details in \S\ref{subsec:eval-alternate}.









\subsection{Overall improvements}
\label{subsec:eval:overall}

We evaluate \name and the baselines along three dimensions---{\em inference accuracy} (\% of images correctly classified), {\em resource consumption} (in GPU time), and {\em capacity} (the number of concurrently processed video streams).
Note that the performance is always tested while keeping up with the original video frame rate (\ie no indefinite frame queueing).

\begin{figure}
\captionsetup[subfigure]{justification=centering}
  \centering
  \begin{subfigure}[t]{0.9\linewidth}
    \centering
    \includegraphics[width=\linewidth]{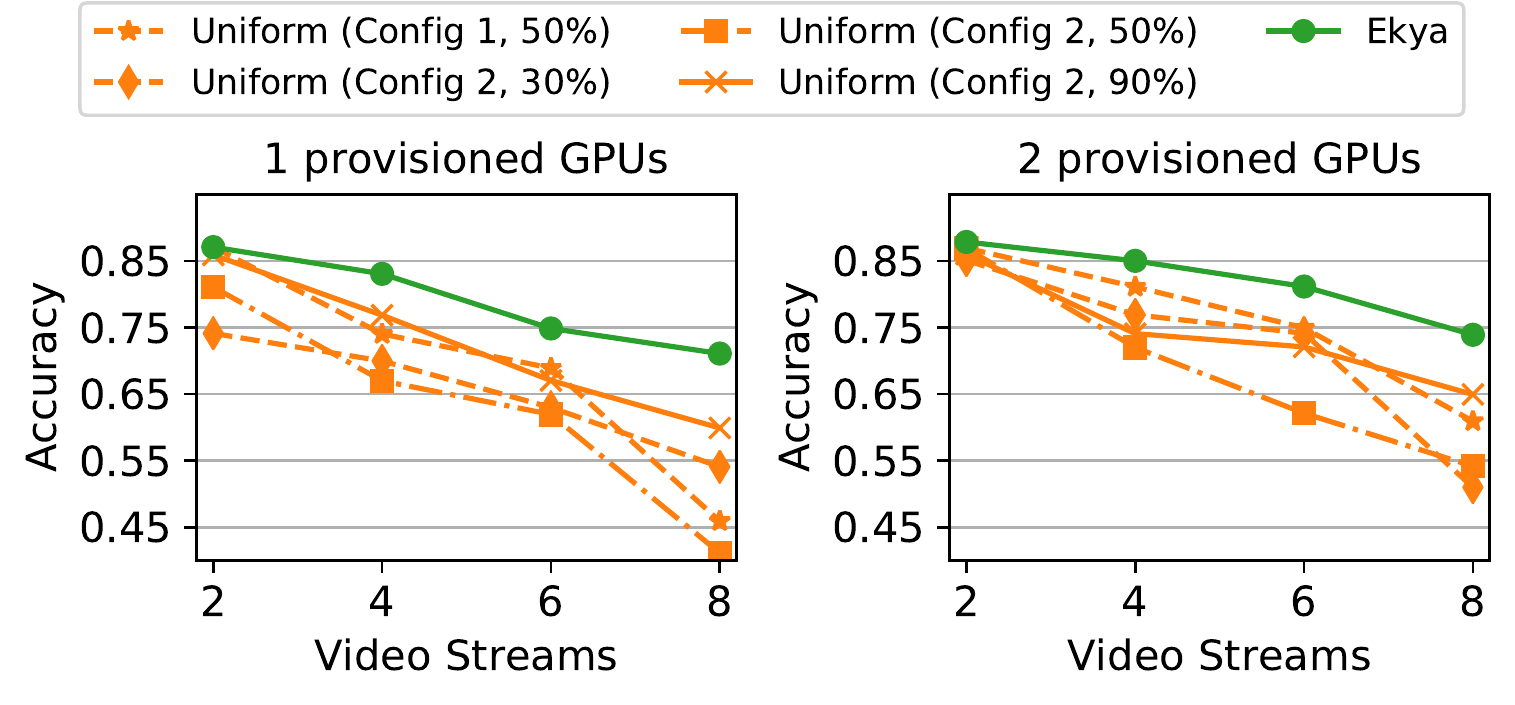}
    \caption{\small Cityscapes}
    \label{fig:sys-impl-cityscapes}
  \end{subfigure}
  \\
  \begin{subfigure}[t]{0.9\linewidth}
    \centering
    \includegraphics[width=\linewidth]{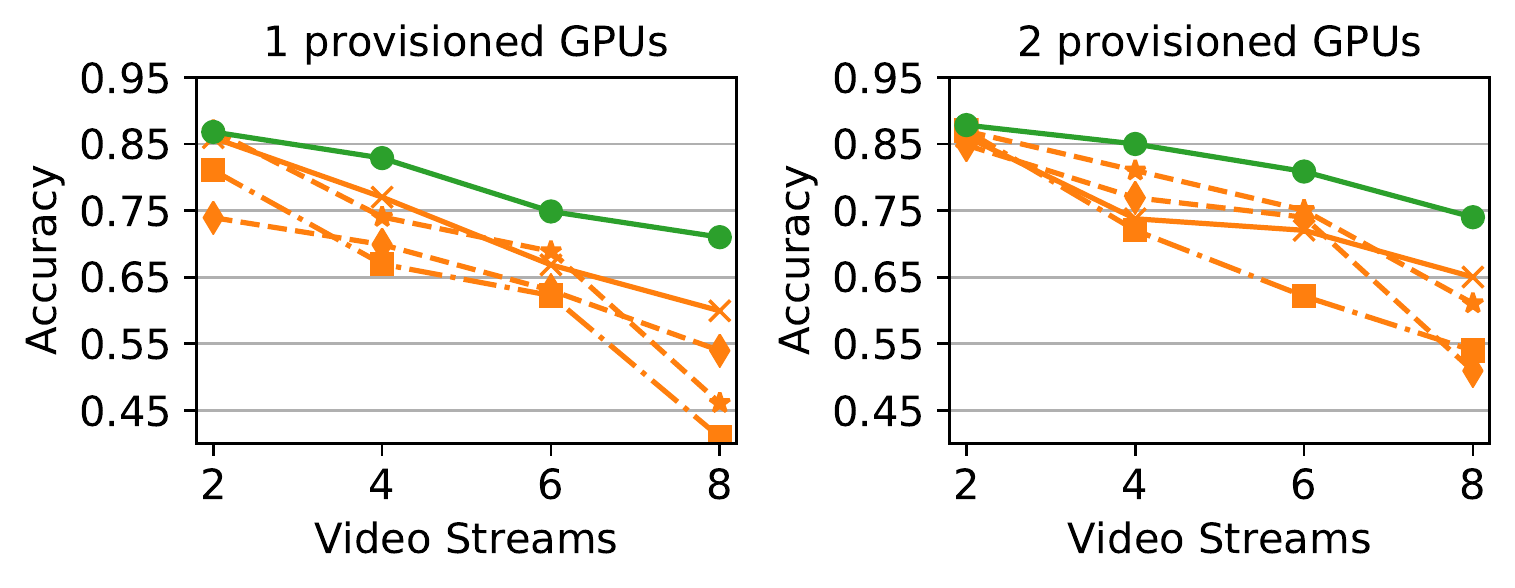}
    \caption{\small Waymo}
    \label{fig:sys-impl-waymo}
  \end{subfigure}
  ~~~
  \caption{\small \bf  Effect of adding more video streams on accuracy for different schedulers. When more video streams sharing the same resource, \name's accuracy gracefully degrades while the \fair baselines' accuracy drops faster. (``Uniform (Config 1, 90\%)'' means when \fair scheduler allocates 90\% GPU resource to inference 10\% to retraining.)
  }
  \label{fig:scalability-sysimpl-fixedGPUs-accuracy}
\end{figure}

\mypara{Accuracy vs. Number of concurrent video streams}
Figure~\ref{fig:scalability-sysimpl-fixedGPUs-accuracy} shows that the accuracy of \name and the \fair baselines when analyzing a growing number of concurrent video streams under a fixed number of provisioned GPUs for Waymo and Cityscapes datasets. 
The \fair baselines use different combinations of pre-determined retraining configurations and resource partitionings.
As the number of video streams increases, \name enjoys a growing advantage (upto 29\% under 1 GPU and 23\% under 2 GPU) in accuracy over the \fair baselines. 
This is because \name gradually shifts more resource from retraining to inference and uses cheaper retraining configurations. 
In contrast, increasing the number of streams forces the \fair baseline to allocate less GPU cycles to each inference job, while retraining jobs, which use fixed configurations, slow down and take the bulk of each window.
This trend persists with different GPUs.

\begin{table}[]
\footnotesize
\begin{tabular}{cccc}
\hline
\multirow{2}{*}{Scheduler} & \multicolumn{2}{c}{Capacity} & \multirow{2}{*}{Scaling factor} \\ \cline{2-3}
& 1 GPU & 2 GPUs &  \\ \hline
\textbf{Ekya} & \textbf{2} & \textbf{8} & \textbf{4x} \\ \hline
Uniform (Config 1, 50\%) & 2 & 2 & 1x \\ \hline
Uniform (Config 2, 90\%) & 2 & 4 & 2x \\ \hline
Uniform (Config 2, 50\%) & 2 & 4 & 2x \\ \hline
Uniform (Config 2, 30\%) & 0 & 2 & - \\ \hline
\end{tabular}
\caption{\small \bf Capacity (number of video streams that can be concurrently supported subject to accuracy target 0.75) vs. number of provisioned GPUs.
\name scales better than the \fair baselines with more available compute resource.
}
\label{tab:scalability-gpu-vs-cam-thresholded}
\end{table}



\mypara{Number of video streams vs. provisioned resource}
We compare \name's {\em capacity} (defined by the maximum number of concurrent video streams subject to an accuracy threshold) with that of \fair baseline, as more GPUs are available.
Setting an accuracy threshold is common in practice, since applications usually require accuracy to be above a threshold for the inference to be usable.
Table~\ref{tab:scalability-gpu-vs-cam-thresholded} uses the Cityscapes results (Figure~\ref{fig:scalability-sysimpl-fixedGPUs-accuracy}) to derive the scaling factor of capacity vs. the number of provisioned GPUs and shows that with more provisioned GPUs, \name scales faster than \fair baselines.

\begin{figure}
  \centering
  \begin{subfigure}[t]{0.49\linewidth}
    \centering
    \includegraphics[width=\linewidth]{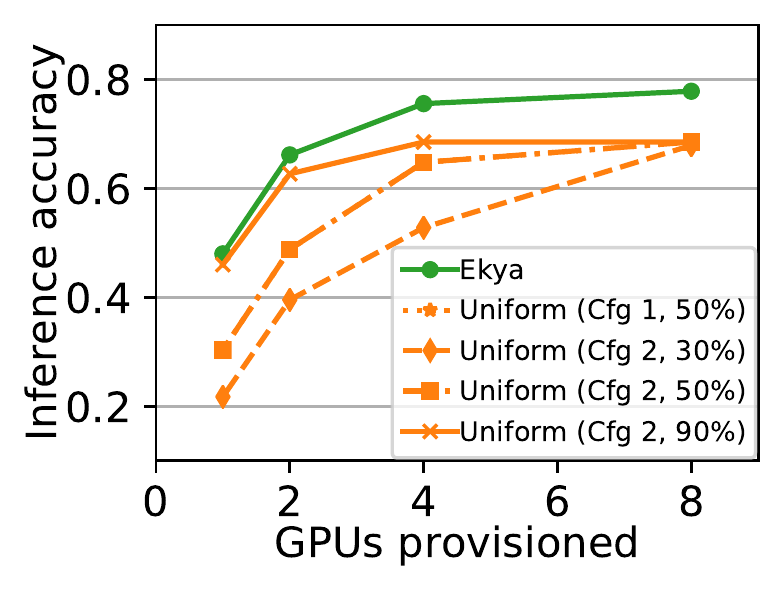}
    \caption{\small Cityscapes}
    \label{fig:scalability-gpus-cityscapes-golden}
  \end{subfigure}
  ~~~
  \begin{subfigure}[t]{0.49\linewidth}
    \centering
    \includegraphics[width=\linewidth]{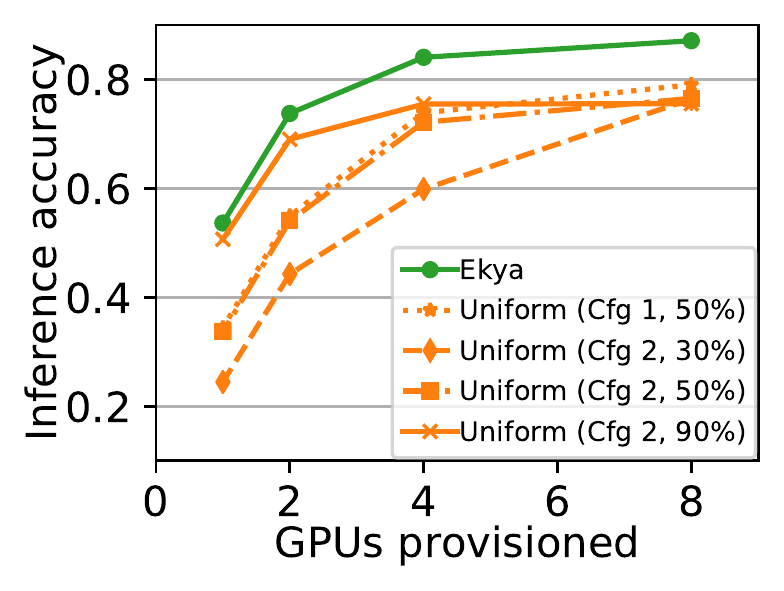} 
    \caption{\small Waymo}
    \label{fig:scalability-gpus-waymo-golden}
  \end{subfigure}
  \\
  \begin{subfigure}[t]{0.49\linewidth}
    \centering
    \includegraphics[width=\linewidth]{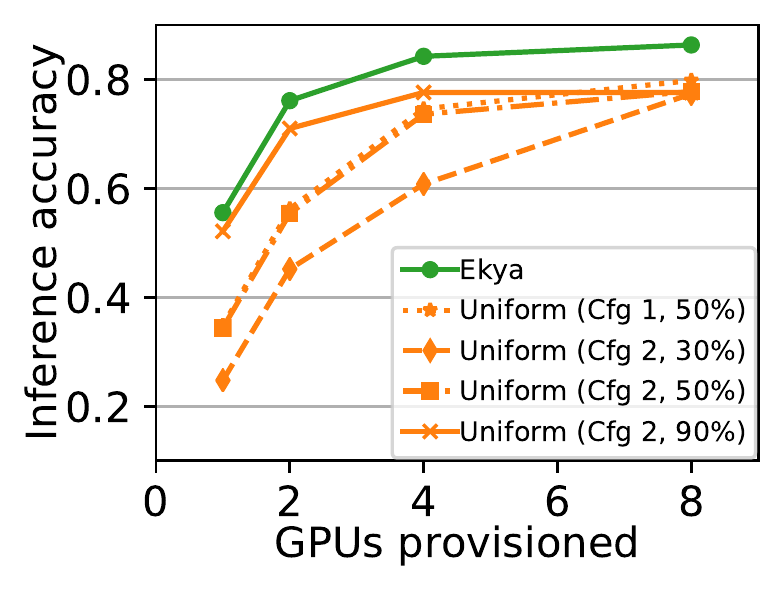} 
    \caption{\small Urban Building}
    \label{fig:scalability-gpus-lasvegas-golden}
  \end{subfigure}
  ~~~
  \begin{subfigure}[t]{0.49\linewidth}
    \centering
    \includegraphics[width=\linewidth]{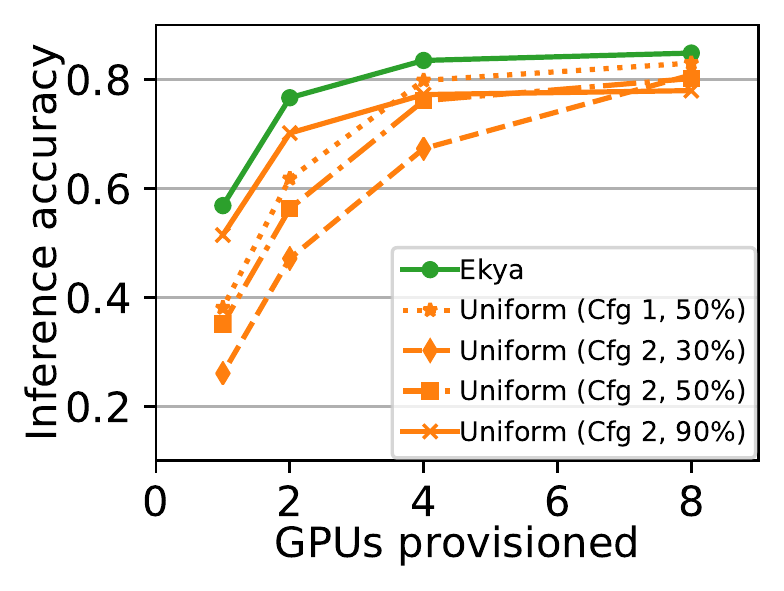} 
    \caption{\small Urban Traffic}
    \label{fig:scalability-gpus-bellevue-golden}
  \end{subfigure}
  \caption{\small \bf Inference accuracy of different schedulers when processing 10 video streams under varying GPU provisionings.
  }
  \label{fig:scalability-gpus}
\end{figure}

\mypara{Accuracy vs. provisioned resource}
Finally, Figure~\ref{fig:scalability-gpus} stress-tests \name and the \fair baselines to process 10 concurrent video streams and shows their average inference accuracy under different number of GPUs. 
To scale to more GPUs, we use the simulator (\S\ref{subsec:eval-setup}), which uses profiles recorded from real tests and we verified that it produced similar results as the implementation at small-scale.
As we increase the number of provisioned GPUs, we see that \name consistently outperforms the best of the two baselines by a considerable margin and more importantly, with 4 GPUs \name achieves higher accuracy (marked with the dotted horizontal line) than the baselines at 16 GPUs (\ie 4$\times$ resource saving).




\mypara{Summary}
The results highlight two advantages of \name. 
First, it allocates resources to retraining only when the accuracy gain from the retraining outweighs the temporary inference accuracy drop due to frame subsampling.
Second, when it allocates resource to retraining, it retrains the model with a configuration that can finish in time for the inference to leverage the higher accuracy from the retrained model. 

\subsection{Understanding Ekya improvements}
\label{subsec:eval-understanding}

\begin{figure} [t!]
 	\includegraphics[width=0.9\linewidth]{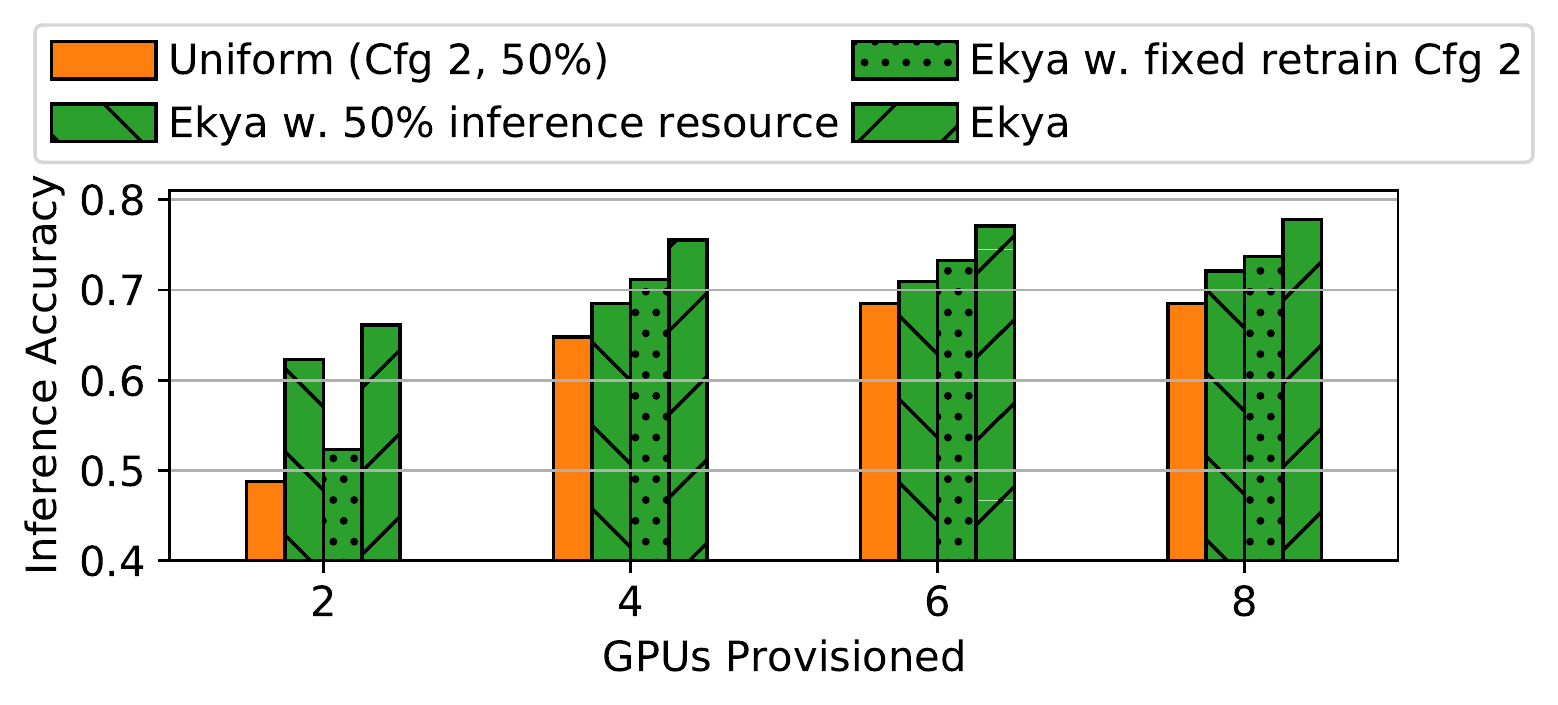}
	\caption{\small \bf A factor analysis of \name that shows the impact of removing dynamic resource allocation (Ekya-FixedRes) or removing retraining configuration adaptation (Ekya-FixedConfig). 
	}
	\label{fig:factor-analysis}
\end{figure}

\mypara{Component-wise contribution}
Figure~\ref{fig:factor-analysis} understands the contributions of resource allocation and configuration selection (on 10 video streams with 4 GPUs provisioned). 
We construct two variants from {\name}:
\emph{Ekya-FixedRes}, which removes the smart resource allocation in \name (\ie using the inference/training resource partition of the \fair baseline), 
and \emph{Ekya-FixedConfig} removes the microprofiling-based configuration selection in \name (\ie using the fixed configuration of the \fair baseline). 
Figure~\ref{fig:factor-analysis} shows that both adaptive resource allocation and configuration selection has a substantial contribution to \name{}'s gains in accuracy, especially when the system is under stress (\ie fewer resources are provisioned). 

\begin{figure}
\captionsetup[subfigure]{justification=centering}
  \centering
  \begin{subfigure}[t]{0.5\linewidth}
    \centering
    \includegraphics[width=\linewidth]{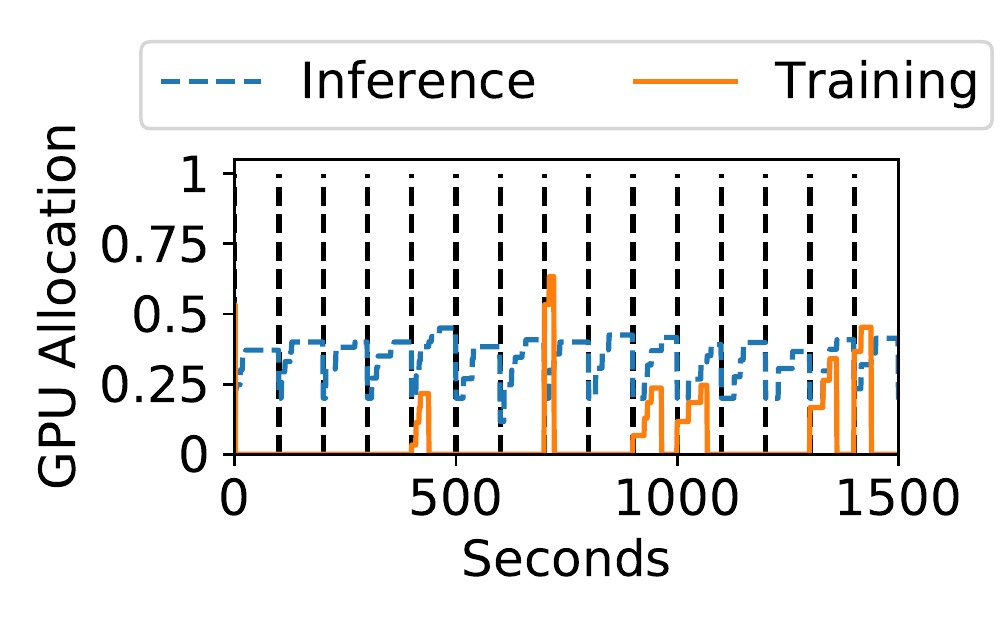}
    \caption{\small Video stream \#1 \\(Inference accuracy = 0.82)}
    \label{fig:temporal-video-1}
  \end{subfigure}
  ~
  \begin{subfigure}[t]{0.5\linewidth}
    \centering
    \includegraphics[width=\linewidth]{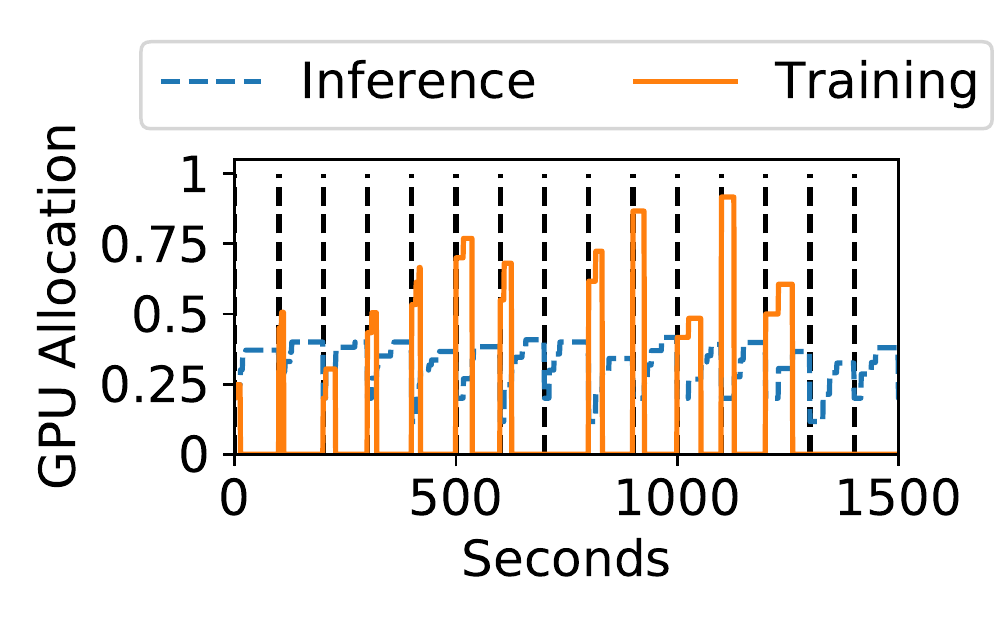} 
    \caption{\small Video stream \#2 \\(Inference accuracy = 0.83)}
    \label{fig:temporal-video-2}
  \end{subfigure}
  \caption{\small \bf Resource allocation of two ``Urban Building'' video streams over retraining windows. Unlike the \fair baseline, \name adapts when to retrain each stream's model and allocates resource based on the retraining benefit to each stream.}
  \label{fig:temporal-resource-allocation}
\end{figure}

\mypara{Resource allocation across streams}
Figure~\ref{fig:temporal-resource-allocation} shows \name's resource allocation across two example video streams over several retraining windows. 
In contrast to the \fair baselines that use the same retraining configuration and allocate equal resource to retraining and inference (when retraining takes place), \name retrains the model only when it benefits and allocates different amounts of GPUs to the retraining jobs of video streams, depending on how much accuracy gain is expected from retraining on each stream. 
In this case, more resource is diverted to video stream \#1 (\#1 can benefit more from retraining than \#2) and both video streams achieve much higher accuracies (0.82 and 0.83) than the \fair baseline.

\begin{figure}
  \centering
  \begin{subfigure}[t]{0.95\linewidth}
    \centering
    \includegraphics[width=\linewidth]{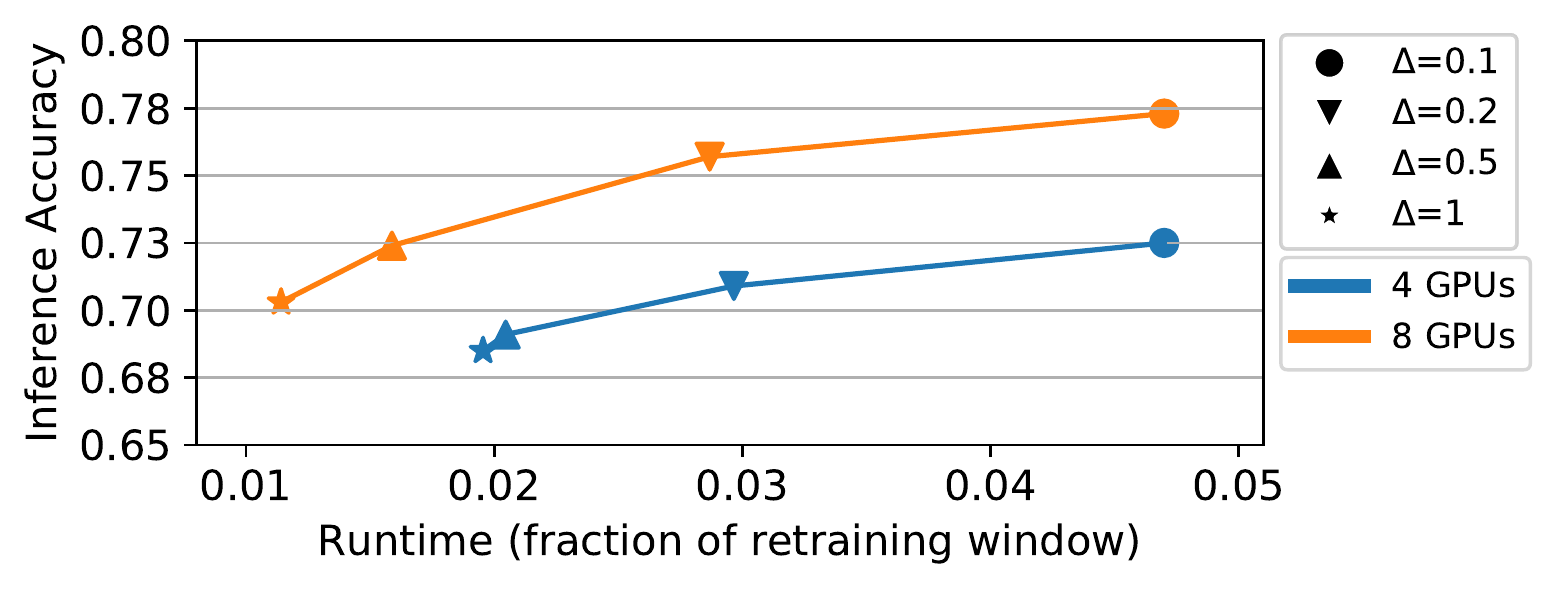}
  \end{subfigure}
  \caption{\small \bf {Effect of the $\Delta$ parameter on the thief scheduler. Smaller values increase the runtime (though still a tiny fraction of a retraining window of 200s) but improve the accuracy.}}
  \label{fig:sensitivity-delta}
\end{figure}

\mypara{Impact of scheduling granularity}
A key parameter in \name's scheduling algorithm (\S\ref{subsec:thief}) is the allocation quantum $\Delta$: it controls the runtime of the scheduling algorithm and the granularity of resource allocation.
Figure~\ref{fig:sensitivity-delta} plots this tradeoff with the same setting as Figure~\ref{fig:scalability-gpus} (10 video streams). 
While increasing $\Delta$ from $1.0$ (coarse-grained; one full GPU) to $0.1$ (fine-grained; fraction of a GPU), we see the accuracy increases substantially $\sim8\%$.
Though the runtime also increases to 9.5 seconds, it is still a tiny fraction ($4.7\%$) of the retraining window ($200$s), and we use $\Delta=0.1$ in our experiments.


\subsection{Effectiveness of micro-profiling}
\label{subsec:eval-profiling}

Finally, we examine the effectiveness of the micro-profiler.


\begin{figure}
\captionsetup[subfigure]{justification=centering}
  \centering
  \begin{subfigure}[t]{0.5\linewidth}
    \centering
    \includegraphics[width=\linewidth]{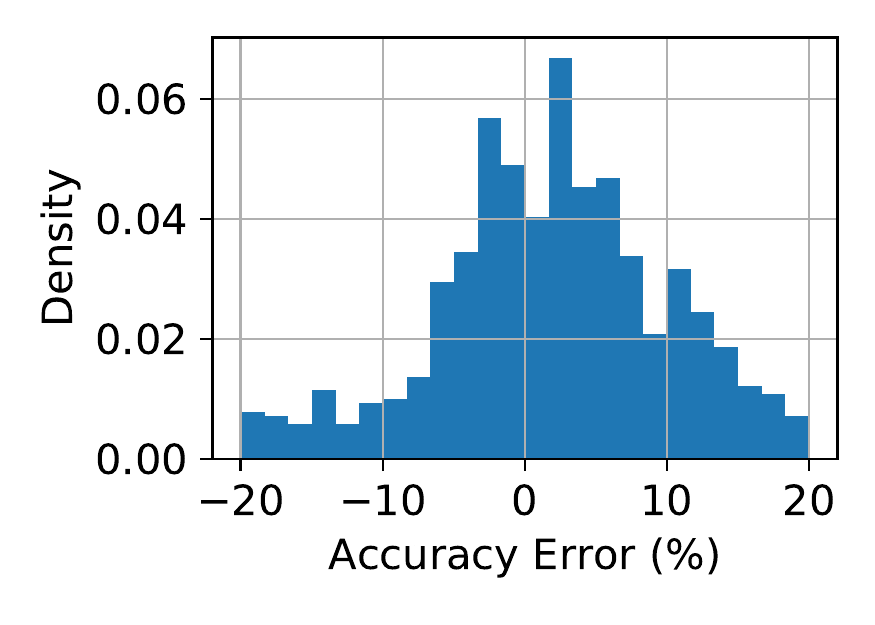}
    \caption{\small Distribution of accuracy estimation errors.}
    \label{fig:microprofiling-benchmark}
  \end{subfigure}
  ~
  \begin{subfigure}[t]{0.5\linewidth}
    \centering
    \includegraphics[width=\linewidth]{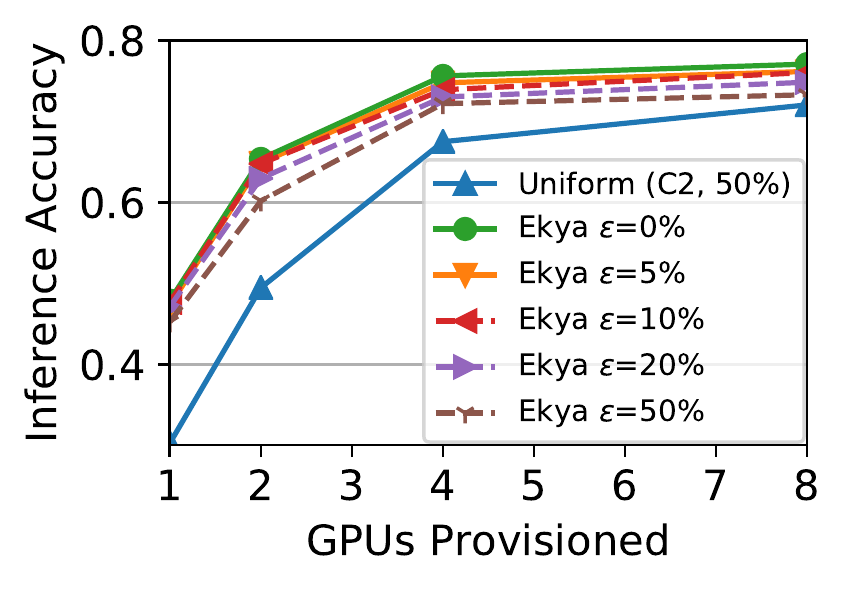} 
    \caption{\small Impact of an controlled error $\epsilon$ to accuracy estimates.}
    \label{fig:sensitivity-accuracy-error}
  \end{subfigure}
  \caption{\small Evaluation of microprofiling performance. (a) shows the distribution of microprofiling's actual estimation errors, and (b) shows the robustness of \name's performance against microprofiling's estimation errors. 
  }
\end{figure}


\mypara{Errors of microprofiled accuracy estimates}
\name's micro-profiler estimates the accuracy of each configuration (\S\ref{subsec:profiling}) by training it on a subset of the data for a small number of epochs. 
To evaluate the micro-profiler's estimates, we run it on all configurations for 5 epochs and on 10\% of the retraining data from all streams of the Cityscapes dataset, and calculate the estimation error with respect to the retrained accuracies when trained on 100\% of the data for 5, 15 and 30 epochs. 
Figure~\ref{fig:microprofiling-benchmark} plots the distribution of the errors in accuracy estimation and and show that the micro-profiled estimates are largely unbiased with an median absolute error of 5.8\%. 

\begin{table}[]
\footnotesize
\begin{tabular}{cccccc}
\hline
\multirow{2}{*}{} & \multicolumn{2}{c}{Bandwidth (Mbps)} & \multirow{2}{*}{Accuracy} & \multicolumn{2}{c}{More bandwidth needed} \\ \cline{2-3} \cline{5-6} 
 & Uplink & Downlink &  & Uplink & Downlink \\ \hline
Cellular & 5.1 & 17.5 & 68.5\% & 10.2$\times$ & 3.8$\times$ \\ \hline
Satellite & 8.5 & 15 & 69.2\% & 5.9$\times$ & 4.4$\times$ \\ \hline
Cellular ($2\times$) & 10.2 & 35 & 71.2\% & 5.1$\times$ & 1.9$\times$ \\ \hline
{\bf \name} & - & - & {\bf 77.8\%} & {\bf -} & {\bf -} \\ \hline
\end{tabular}
\caption{\label{tab:cloudexpt}\small\bf Retraining in the cloud under different networks~\cite{39-getmobile, 57-getmobile, getmobile} versus using {\name} at the edge. \name achieves better accuracy without using expensive satellite and cellular links. 
}
\end{table}

\mypara{Sensitivity to microprofiling estimation errors}
Finally, we test the impact of accuracy estimation errors (\S\ref{subsec:profiling}) on \name's inference accuracy.
We add a controlled Gaussian noise on top of the real retraining accuracy as the predictions when the microprofiler is queried. 
Figure~\ref{fig:sensitivity-accuracy-error} shows that \name is quite robust to accuracy estimate errors: with upto 20\% errors (in which all errors in Figure~\ref{fig:microprofiling-benchmark} lie) in the profiler prediction, the maximum accuracy drop is merely 3\%. 



\subsection{Comparison with alternative designs}
\label{subsec:eval-alternate}

\mypara{Ekya vs. Cloud-based retraining}
One may upload a sub-sampled video stream to the cloud, retrain the model in the cloud, and download the model to the edge server~\cite{khani2020real}. 
While this solution compromises the privacy of the videos and is not an option for many deployments due to legal stipulations \cite{sweden-data, azure-data}, we nonetheless evaluate this option as it lets the edge servers focus on inference. We conclude that the cloud-based solution results in lower accuracy due to significant network delays 
on the constrained networks typical of edges \cite{getmobile}.

As a simple example, consider 8 video streams with a ResNet18 model and a retraining window of 400 seconds. 
For a HD (720p) video stream at 4Mbps and 10\% data sub-sampling (typical in our experiments), this amounts to 160Mb of training data per camera per window. 
Uploading 160Mb for each of the 8 cameras over a 4G cellular uplink (5.1 Mbps \cite{57-getmobile}) and downloading the trained ResNet18 models (each of size of 398 Mb~\cite{torchvision-models}) over the 17.5 Mbps downlink \cite{57-getmobile} takes a total of 432 seconds (even excluding the model retraining time), which already exceeds the retraining window.


To test on the Cityscapes dataset, we extend our simulator (\S\ref{subsec:eval-setup}) to account for network delays during retraining, and test with 8 videos and 4 GPUs. We use the conservative assumption that retraining in the cloud is ``instantaneous'' (cloud GPUs are powerful than edge GPUs). Table \ref{tab:cloudexpt} lists the accuracies with cellular 4G links (one and two subscriptions) and a satellite link, which are both indicative of edge deployments \cite{getmobile}. We use two cellular links so as to meet the 400s retraining window (based on the description above). 

For the cloud alternatives to match {\name}'s accuracy, we will need to provision additional uplink capacity of 5$\times$-10$\times$ and downlink capacity of 2$\times$-4$\times$ (of the already expensive links). In summary, {\name}'s edge-based solution is better than a cloud based alternate for retraining in {\em both metrics of} accuracy as well as network usage (\name sends no data out of the edge), all while providing privacy for the videos. 

\mypara{Ekya vs. Re-using pretrained models}
Another alternative to \name's continuous retraining is to cache retrained models and reuse them, \eg pick the model that was trained on a similar class distribution. 
To test this baseline, we pre-train and cache a few tens of DNNs from earlier retraining windows from the Cityscapes dataset. 
In each retraining window of our experiment with 8 GPUs and 10 video streams, we pick the cached DNN whose class distribution (vector of object class frequencies) of its training data has the closest Euclidean distance with the current window's data. GPU cycles are evenly shared by the inference jobs (since there is no retraining).
The resulting average inference accuracy is 0.72 is lower than the \name's accuracy of 0.78 (see Figure~\ref{fig:scalability-gpus-cityscapes-golden}).
This is because even though the class distributions may be similar, the models cannot be directly reused from any window as the appearances of objects may still differ considerably (Figure \ref{fig:cityscapes-motivation}).


\section{Related Work}
\label{sec:related_work}








\noindent\textbf{1) ML training systems.} 
For large scale training in the cloud, model and data parallel frameworks \cite{DBLP:conf/nips/DeanCMCDLMRSTYN12, DBLP:journals/pvldb/LowGKBGH12, 199317, mxnet} 
and various resource schedulers \cite{DBLP:journals/corr/abs-1907-01484, DBLP:conf/osdi/XiaoBRSKHPPZZYZ18, DBLP:conf/nsdi/GuCSZJQLG19, DBLP:conf/eurosys/PengBCWG18, DBLP:conf/sigcomm/GrandlAKRA14, DBLP:conf/cloud/ZhangSOF17} 
have been developed to schedule the cluster for ML workloads. These systems, however, target different objectives than {\name}, like maximizing parallelism, efficiency, fairness, or minimizing average job completion. 
Collaborative training systems \cite{DBLP:journals/corr/abs-1902-01046, DBLP:conf/edge/LuSTLZCP19} work on decentralized data on mobile phones. Their focus is on coordinating the training between phones and the cloud, and not on training the models alongside inference.

\noindent\textbf{2) Video analytics systems.} Prior work has built low-cost, high-accuracy and highly scalable video analytics systems across the edge and cloud~\cite{videostorm,chameleon,noscope}. VideoStorm~\cite{videostorm} investigates quality-lag requirements in video queries. NoScope exploits difference detectors and cascaded models to speedup queries~\cite{noscope}. Focus uses low-cost models to index videos~\cite{DBLP:conf/osdi/HsiehABVBPGM18}. Chameleon exploits correlations in camera content to amortize {\em profiling costs}~\cite{chameleon}. All of these works optimize only the inference accuracy unlike \name's focus on retraining.

\noindent\textbf{3) Hyper-parameter optimization.} 
Efficient techniques to explore the space of hyper-parameters is crucial in training systems to find the model with the best accuracy. Techniques range from simplistic grid or random search~\cite{DBLP:journals/jmlr/BergstraB12}, to more sophisticated approaches using random forests~\cite{DBLP:conf/lion/HutterHL11}, Bayesian optimization~\cite{DBLP:conf/nips/SnoekLA12, Swersky_scalablebayesian}, probabilistic model~\cite{DBLP:conf/middleware/RasleyH0RF17}, or non-stochastic infinite-armed bandits~\cite{DBLP:journals/jmlr/LiJDRT17}. Unlike the focus of these techniques towards finding the hyper-parameters that train the model with the highest accuracy, our focus is on resource allocation. Further, we are focused on the inference accuracy over the retrained window, where often producing the best retrained model turns out to be sub-optimal.

\noindent\textbf{4) Continuous learning.} Machine learning literature on continuous learning adapts models as new data comes in. 
A common approach used is transfer learning~\cite{DBLP:journals/corr/RazavianASC14,DBLP:conf/edge/LuSTLZCP19,mullapudi2019,44873}. Research has also been conducted on handling catastrophic forgetting~\cite{DBLP:journals/corr/abs-1708-01547,datadrift-a}, using limited amount of training data~\cite{icarl-14,DBLP:journals/corr/abs-1905-10887}, and dealing with class imbalance~\cite{Belouadah_2019_ICCV,DBLP:journals/corr/abs-1905-13260}. 
{\name} builds atop continuous learning techniques for its scheduling and implementation, to enable them in edge deployments. 

\noindent\textbf{5) Edge compute systems.} By deploying computation close to the data sources, edge computing benefits many applications, including video analytics \cite{edgevideo-1, ieee-computer, getmobile}. While there has been edge-based solutions for video processing~\cite{videoedge}, we enable joint optimization of video inference and retraining.

\section{Conclusion}
\label{sec:conclusion}


Continuous learning enables edge DNNs to maintain high accuracy even with data drift, but it also poses a complex and fundamental tradeoff between retraining and inference. We introduce {\name}, an efficient system that maximizes inference accuracy by balancing across multiple retraining and inference tasks. 
{\name}'s resource scheduler makes the problem practical and tractable by pruning the large decision space and prioritizing promising retraining tasks. {\name}'s performance estimator provides essential accuracy estimation with very little overheads. Our evaluation with a diverse set of of video streams shows that {\name} achieves $29\%$ higher accuracy than a baseline scheduler, and the baseline needs $4\times$ more GPU resources to achieve {\name}'s accuracy. We conclude {\name} is a practical system for continuous learning for video analytics on the edge, and we hope that our findings will spur further research into the tradeoff between retraining and inference.

\bibliographystyle{abbrv}
\bibliography{main}

\begin{thebibliography}{10}

\bibitem{mxnet}
{MxNet: a flexible and efficient library for deep learning}.
\newblock \url{https://mxnet.apache.org/}.

\bibitem{nvidia-mps}
Nvidia multi-process service.
\newblock
  \url{https://docs.nvidia.com/deploy/pdf/CUDA_Multi_Process_Service_Overview.pdf}.
\newblock (Accessed on 09/16/2020).

\bibitem{nnls}
scipy.optimize.nnls — scipy v1.5.2 reference guide.
\newblock
  \url{https://docs.scipy.org/doc/scipy/reference/generated/scipy.optimize.nnls.html}.
\newblock (Accessed on 09/17/2020).

\bibitem{torch-checkpoint}
torch.utils.checkpoint — pytorch 1.6.0 documentation.
\newblock \url{https://pytorch.org/docs/stable/checkpoint.html}.
\newblock (Accessed on 09/16/2020).

\bibitem{torchvision-models}
torchvision.models — pytorch 1.6.0 documentation.
\newblock \url{https://pytorch.org/docs/stable/torchvision/models.html}.
\newblock (Accessed on 09/16/2020).

\bibitem{hyperparameter-16}
{A Comprehensive List of Hyperparameter Optimization \& Tuning Solutions}.
\newblock
  \url{https://medium.com/@mikkokotila/a-comprehensive-list-of-hyperparameter-optimization-tuning-solutions-88e067f19d9},
  2018.

\bibitem{fair-2}
{A. Ghodsi, M. Zaharia, B. Hindman, A. Konwinski, S. Shenker, and I. Stoica}.
\newblock Fair allocation of multiple resource types.
\newblock In {\em USENIX NSDI}, 2011.

\bibitem{199317}
M.~Abadi, P.~Barham, J.~Chen, Z.~Chen, A.~Davis, J.~Dean, M.~Devin,
  S.~Ghemawat, G.~Irving, M.~Isard, M.~Kudlur, J.~Levenberg, R.~Monga,
  S.~Moore, D.~G. Murray, B.~Steiner, P.~Tucker, V.~Vasudevan, P.~Warden,
  M.~Wicke, Y.~Yu, and X.~Zheng.
\newblock {TensorFlow: A System for Large-Scale Machine Learning}.
\newblock In {\em {USENIX} OSDI}, 2016.

\bibitem{azure-data}
{Achieving Compliant Data Residency and Security with Azure}.

\bibitem{openai-blog}
{AI and Compute}.
\newblock {https://openai.com/blog/ai-and-compute/}, 2018.

\bibitem{ieee-computer}
G.~Ananthanarayanan, V.~Bahl, P.~Bodík, K.~Chintalapudi, M.~Philipose, L.~R.
  Sivalingam, and S.~Sinha.
\newblock {Real-time Video Analytics – the killer app for edge computing}.
\newblock {\em IEEE Computer}, 2017.

\bibitem{aws-outposts}
{AWS Outposts}.
\newblock {https://aws.amazon.com/outposts/}.

\bibitem{azure-ase}
{Azure Stack Edge}.
\newblock {https://azure.microsoft.com/en-us/services/databox/edge/}.

\bibitem{Belouadah_2019_ICCV}
E.~Belouadah and A.~Popescu.
\newblock {IL2M: Class Incremental Learning With Dual Memory}.
\newblock In {\em IEEE ICCV}, 2019.

\bibitem{DBLP:journals/jmlr/BergstraB12}
J.~Bergstra and Y.~Bengio.
\newblock {Random Search for Hyper-Parameter Optimization}.
\newblock {\em J. Mach. Learn. Res.}, 13:281--305, 2012.

\bibitem{DBLP:journals/corr/abs-1902-01046}
K.~Bonawitz, H.~Eichner, W.~Grieskamp, D.~Huba, A.~Ingerman, V.~Ivanov,
  C.~Kiddon, J.~Konecn{\'{y}}, S.~Mazzocchi, H.~B. McMahan, T.~V. Overveldt,
  D.~Petrou, D.~Ramage, and J.~Roselander.
\newblock {Towards Federated Learning at Scale: System Design}.
\newblock In {\em SysML}, 2019.

\bibitem{kubernetes}
B.~Burns, B.~Grant, D.~Oppenheimer, E.~Brewer, and J.~Wilkes.
\newblock Borg, omega, and kubernetes.
\newblock {\em ACM Queue}, 14:70--93, 2016.

\bibitem{videoedge}
{Chien-Chun Hung, Ganesh Ananthanarayanan, Peter Bodik, Leana Golubchik, Minlan
  Yu, Paramvir Bahl, Matthai Philipose}.
\newblock Videoedge: Processing camera streams using hierarchical clusters.
\newblock In {\em ACM/IEEE SEC}, 2018.

\bibitem{37-getmobile}
{CLIFFORD, M. J., PERRONS, R. K., ALI, S. H.,ANDGRICE, T. A.}
\newblock {Extracting Innovations: Mining, Energy, and Technological Changein
  the Digital Age}.
\newblock In {\em CRC Press}, 2018.

\bibitem{cnn-perf}
{cnn-benchmarks}.
\newblock {https://github.com/jcjohnson/cnn-benchmarks\#resnet-101}, {2017}.

\bibitem{noscope}
{D. Kang, J. Emmons, F. Abuzaid, P. Bailis and M. Zaharia}.
\newblock Noscope: Optimizing neural network queries over video at scale.
\newblock In {\em VLDB}, 2017.

\bibitem{datadrift-b}
{D Maltoni, V Lomonaco}.
\newblock Continuous learning in single-incremental-task scenarios.
\newblock In {\em Neural Networks}, 2019.

\bibitem{DBLP:conf/nips/DeanCMCDLMRSTYN12}
J.~Dean, G.~Corrado, R.~Monga, K.~Chen, M.~Devin, Q.~V. Le, M.~Z. Mao,
  M.~Ranzato, A.~W. Senior, P.~A. Tucker, K.~Yang, and A.~Y. Ng.
\newblock {Large Scale Distributed Deep Networks}.
\newblock In {\em NeurIPS}, 2012.

\bibitem{chick-fill}
{Edge Computing at Chick-fil-A}.
\newblock
  {https://medium.com/@cfatechblog/edge-computing-at-chick-fil-a-7d67242675e2}.
\newblock 2019.

\bibitem{bellevue-report}
{Ganesh Ananthanarayanan, Victor Bahl, Yuanchao Shu, Franz Loewenherz, Daniel
  Lai, Darcy Akers, Peiwei Cao, Fan Xia, Jiangbo Zhang, Ashley Song}.
\newblock {Traffic Video Analytics – Case Study Report}.
\newblock 2019.

\bibitem{continuous-12}
{GI Parisi, R Kemker, JL Part, C Kanan, S Wermter }.
\newblock Continual lifelong learning with neural networks: A review.
\newblock In {\em Neural Networks}, 2019.

\bibitem{vizier}
D.~Golovin, B.~Solnik, S.~Moitra, G.~Kochanski, J.~Karro, and D.~Sculley.
\newblock Google vizier: A service for black-box optimization.
\newblock In {\em Proceedings of the 23rd ACM SIGKDD International Conference
  on Knowledge Discovery and Data Mining}, KDD '17, page 1487–1495, New York,
  NY, USA, 2017. Association for Computing Machinery.

\bibitem{tetris}
R.~Grandl, G.~Ananthanarayanan, S.~Kandula, S.~Rao, and A.~Akella.
\newblock Multi-resource packing for cluster schedulers.
\newblock In {\em Proceedings of the 2014 ACM Conference on SIGCOMM}, SIGCOMM
  '14, page 455–466, New York, NY, USA, 2014. Association for Computing
  Machinery.

\bibitem{DBLP:conf/sigcomm/GrandlAKRA14}
R.~Grandl, G.~Ananthanarayanan, S.~Kandula, S.~Rao, and A.~Akella.
\newblock Multi-resource packing for cluster schedulers.
\newblock In {\em {ACM} {SIGCOMM}}, 2014.

\bibitem{DBLP:conf/nsdi/GuCSZJQLG19}
J.~Gu, M.~Chowdhury, K.~G. Shin, Y.~Zhu, M.~Jeon, J.~Qian, H.~H. Liu, and
  C.~Guo.
\newblock Tiresias: {A} {GPU} cluster manager for distributed deep learning.
\newblock In {\em {USENIX} {NSDI}}, 2019.

\bibitem{videostorm}
{Haoyu Zhang, Ganesh Ananthanarayanan, Peter Bodík, Matthai Philipose, Victor
  Bahl, Michael Freedman}.
\newblock Live video analytics at scale with approximation and delay-tolerance.
\newblock In {\em USENIX NSDI}, 2017.

\bibitem{44873}
G.~Hinton, O.~Vinyals, and J.~Dean.
\newblock {Distilling the Knowledge in a Neural Network}.
\newblock In {\em NeurIPS Deep Learning and Representation Learning Workshop},
  2015.

\bibitem{DBLP:conf/osdi/HsiehABVBPGM18}
K.~Hsieh, G.~Ananthanarayanan, P.~Bod{\'{\i}}k, S.~Venkataraman, P.~Bahl,
  M.~Philipose, P.~B. Gibbons, and O.~Mutlu.
\newblock {Focus: Querying Large Video Datasets with Low Latency and Low Cost}.
\newblock In {\em USENIX OSDI}, 2018.

\bibitem{DBLP:conf/lion/HutterHL11}
F.~Hutter, H.~H. Hoos, and K.~Leyton{-}Brown.
\newblock {Sequential Model-Based Optimization for General Algorithm
  Configuration}.
\newblock In {\em Learning and Intelligent Optimization}, 2011.

\bibitem{yolo9000-1}
{Joseph Redmon, Ali Farhadi }.
\newblock Yolo9000: Better, faster, stronger.
\newblock In {\em CVPR}, 2017.

\bibitem{chameleon}
{Junchen Jiang, Ganesh Ananthanarayanan, Peter Bodík, Siddhartha Sen, Ion
  Stoica}.
\newblock Chameleon: Scalable adaptation of video analytics.
\newblock In {\em ACM SIGCOMM}, 2018.

\bibitem{edgevideo-1}
{Junjue Wang, Ziqiang Feng, Shilpa George, Roger Iyengar, Pillai Padmanabhan,
  Mahadev Satyanarayanan}.
\newblock Towards scalable edge-native applications.
\newblock In {\em ACM/IEEE Symposium on Edge Computing}, 2019.

\bibitem{deepresidual-2}
{K He, X Zhang, S Ren, J Sun }.
\newblock Deep residual learning for image recognition.
\newblock In {\em CVPR}, 2016.

\bibitem{khani2020real}
M.~Khani, P.~Hamadanian, A.~Nasr-Esfahany, and M.~Alizadeh.
\newblock Real-time video inference on edge devices via adaptive model
  streaming.
\newblock {\em arXiv preprint arXiv:2006.06628}, 2020.

\bibitem{incremental-15}
{Konstantin Shmelkov, Cordelia Schmid, Karteek Alahari }.
\newblock Incremental learning of object detectors without catastrophic
  forgetting.
\newblock In {\em ICCV}, 2017.

\bibitem{DBLP:journals/corr/abs-1708-01547}
J.~Lee, J.~Yoon, E.~Yang, and S.~J. Hwang.
\newblock {Lifelong Learning with Dynamically Expandable Networks}.
\newblock In {\em ICLR}, 2018.

\bibitem{pbt}
A.~Li, O.~Spyra, S.~Perel, V.~Dalibard, M.~Jaderberg, C.~Gu, D.~Budden,
  T.~Harley, and P.~Gupta.
\newblock A generalized framework for population based training.
\newblock In {\em Proceedings of the 25th ACM SIGKDD International Conference
  on Knowledge Discovery and Data Mining}, KDD '19, page 1791–1799, New York,
  NY, USA, 2019. Association for Computing Machinery.

\bibitem{hyperband}
L.~Li, K.~Jamieson, G.~DeSalvo, A.~Rostamizadeh, and A.~Talwalkar.
\newblock Hyperband: A novel bandit-based approach to hyperparameter
  optimization.
\newblock {\em J. Mach. Learn. Res.}, 18(1):6765–6816, Jan. 2017.

\bibitem{DBLP:journals/jmlr/LiJDRT17}
L.~Li, K.~G. Jamieson, G.~DeSalvo, A.~Rostamizadeh, and A.~Talwalkar.
\newblock Hyperband: {A} novel bandit-based approach to hyperparameter
  optimization.
\newblock {\em J. Mach. Learn. Res.}, 18:185:1--185:52, 2017.

\bibitem{DBLP:journals/pvldb/LowGKBGH12}
Y.~Low, J.~Gonzalez, A.~Kyrola, D.~Bickson, C.~Guestrin, and J.~M. Hellerstein.
\newblock Distributed graphlab: {A} framework for machine learning in the
  cloud.
\newblock {\em {PVLDB}}, 5(8):716--727, 2012.

\bibitem{DBLP:conf/edge/LuSTLZCP19}
Y.~Lu, Y.~Shu, X.~Tan, Y.~Liu, M.~Zhou, Q.~Chen, and D.~Pei.
\newblock Collaborative learning between cloud and end devices: an empirical
  study on location prediction.
\newblock In {\em {ACM/IEEE} {SEC}}, 2019.

\bibitem{datadrift-7}
{M McCloskey, NJ Cohen}.
\newblock Catastrophic interference in connectionist networks: The sequential
  learning problem.
\newblock In {\em Psychology of learning and motivation}, 1989.

\bibitem{compression-17}
{M Sandler, A Howard, Menglong Zhu, Andrey Zhmoginov, Liang-Chieh Chen }.
\newblock Mobilenetv2: Inverted residuals and linear bottlenecks.
\newblock In {\em CVPR}, 2018.

\bibitem{DBLP:journals/mor/MagazineC84}
M.~J. Magazine and M.~Chern.
\newblock A note on approximation schemes for multidimensional knapsack
  problems.
\newblock {\em Math. Oper. Res.}, 9(2), 1984.

\bibitem{themis}
K.~Mahajan, A.~Balasubramanian, A.~Singhvi, S.~Venkataraman, A.~Akella,
  A.~Phanishayee, and S.~Chawla.
\newblock Themis: Fair and efficient {GPU} cluster scheduling.
\newblock In {\em 17th {USENIX} Symposium on Networked Systems Design and
  Implementation ({NSDI} 20)}, pages 289--304, Santa Clara, CA, Feb. 2020.
  {USENIX} Association.

\bibitem{DBLP:journals/corr/abs-1907-01484}
K.~Mahajan, A.~Singhvi, A.~Balasubramanian, S.~Venkataraman, A.~Akella,
  A.~Phanishayee, and S.~Chawla.
\newblock Themis: Fair and efficient {GPU} cluster scheduling for machine
  learning workloads.
\newblock In {\em USENIX NSDI}, 2020.

\bibitem{cityscapes}
{Marius Cordts, Mohamed Omran, Sebastian Ramos, Timo Rehfeld, Markus Enzweiler,
  Rodrigo Benenson, Uwe Franke, Stefan Roth, and Bernt Schiele }.
\newblock The cityscapes dataset for semantic urban scene understanding.
\newblock In {\em CVPR}, 2016.

\bibitem{39-getmobile}
{Measuring Fixed Broadband - Eighth Report, FEDERAL COMMUNICATIONS COMMISSION
  OFFICE OF ENGINEERING AND TECHNOLOGY}.
\newblock
  {https://www.fcc.gov/reports-research/reports/measuring-broadband-america/measuring-fixed-broadband-eighth-report}.
\newblock 2018.

\bibitem{rocket-github}
{Microsoft-Rocket-Video-Analytics-Platform}.
\newblock
  {https://github.com/microsoft/Microsoft-Rocket-Video-Analytics-Platform}.

\bibitem{compression-19}
{Mingxing Tan, Bo Chen, Ruoming Pang, Vijay Vasudevan, Mark Sandler, Andrew
  Howard, Quoc V. Le}.
\newblock Mnasnet: Platform-aware neural architecture search for mobile.
\newblock In {\em CVPR}, 2019.

\bibitem{efficientnet-3}
{Mingxing Tan, Quoc V. Le }.
\newblock Efficientnet: Rethinking model scaling for convolutional neural
  networks.
\newblock In {\em ICML}, 2019.

\bibitem{ray}
P.~Moritz, R.~Nishihara, S.~Wang, A.~Tumanov, R.~Liaw, E.~Liang, M.~Elibol,
  Z.~Yang, W.~Paul, M.~I. Jordan, and I.~Stoica.
\newblock Ray: A distributed framework for emerging ai applications.
\newblock In {\em Proceedings of the 13th USENIX Conference on Operating
  Systems Design and Implementation}, OSDI'18, page 561–577, USA, 2018.
  USENIX Association.

\bibitem{compression-18}
{Ningning Ma, Xiangyu Zhang, Hai-Tao Zheng, and Jian Sun }.
\newblock Shufflenet v2: Practical guidelines for efficient cnn architecture
  design.
\newblock In {\em ECCV}, 2018.

\bibitem{57-getmobile}
{OPENSIGNAL. Mobile Network Experience Report }.
\newblock
  {https://www.opensignal.com/reports/2019/01/usa/mobile-network-experience}.
\newblock 2019.

\bibitem{pytorch}
A.~Paszke, S.~Gross, F.~Massa, A.~Lerer, J.~Bradbury, G.~Chanan, T.~Killeen,
  Z.~Lin, N.~Gimelshein, L.~Antiga, A.~Desmaison, A.~Kopf, E.~Yang, Z.~DeVito,
  M.~Raison, A.~Tejani, S.~Chilamkurthy, B.~Steiner, L.~Fang, J.~Bai, and
  S.~Chintala.
\newblock Pytorch: An imperative style, high-performance deep learning library.
\newblock In H.~Wallach, H.~Larochelle, A.~Beygelzimer, F.~d\textquotesingle
  Alch\'{e}-Buc, E.~Fox, and R.~Garnett, editors, {\em Advances in Neural
  Information Processing Systems 32}, pages 8024--8035. Curran Associates,
  Inc., 2019.

\bibitem{compression-6}
{Pavlo Molchanov, Stephen Tyree, Tero Karras, Timo Aila, Jan Kautz}.
\newblock Pruning convolutional neural networks for resource efficient
  inference.
\newblock In {\em ICLR}, 2017.

\bibitem{waymo}
{Pei Sun and Henrik Kretzschmar and Xerxes Dotiwalla and Aurelien Chouard and
  Vijaysai Patnaik and Paul Tsui and James Guo and Yin Zhou and Yuning Chai and
  Benjamin Caine and Vijay Vasudevan and Wei Han and Jiquan Ngiam and Hang Zhao
  and Aleksei Timofeev and Scott Ettinger and Maxim Krivokon and Amy Gao and
  Aditya Joshi and Yu Zhang and Jonathon Shlens and Zhifeng Chen and Dragomir
  Anguelov}.
\newblock Scalability in perception for autonomous driving: Waymo open dataset,
  2019.

\bibitem{DBLP:conf/eurosys/PengBCWG18}
Y.~Peng, Y.~Bao, Y.~Chen, C.~Wu, and C.~Guo.
\newblock Optimus: an efficient dynamic resource scheduler for deep learning
  clusters.
\newblock In {\em ACM EuroSys}, 2018.

\bibitem{optimus}
Y.~Peng, Y.~Bao, Y.~Chen, C.~Wu, and C.~Guo.
\newblock Optimus: An efficient dynamic resource scheduler for deep learning
  clusters.
\newblock In {\em Proceedings of the Thirteenth EuroSys Conference}, EuroSys
  '18, New York, NY, USA, 2018. Association for Computing Machinery.

\bibitem{DBLP:conf/middleware/RasleyH0RF17}
J.~Rasley, Y.~He, F.~Yan, O.~Ruwase, and R.~Fonseca.
\newblock {HyperDrive: exploring hyperparameters with {POP} scheduling}.
\newblock In {\em {ACM/IFIP/USENIX} Middleware}, 2017.

\bibitem{mullapudi2019}
{Ravi Teja Mullapudi, Steven Chen, Keyi Zhang, Deva Ramanan, Kayvon
  Fatahalian}.
\newblock Online model distillation for efficient video inference.
\newblock In {\em ICCV}, 2019.

\bibitem{DBLP:journals/corr/abs-1905-10887}
S.~V. Ravuri and O.~Vinyals.
\newblock Classification accuracy score for conditional generative models.
\newblock 2019.

\bibitem{DBLP:journals/corr/RazavianASC14}
A.~S. Razavian, H.~Azizpour, J.~Sullivan, and S.~Carlsson.
\newblock {CNN} features off-the-shelf: an astounding baseline for recognition.
\newblock In {\em IEEE CVPR Workshop}, 2014.

\bibitem{20-getmobile}
{Residential landline and fixed broadband services }.
\newblock
  {https://www.ofcom.org.uk/\_\_data/assets/pdf\_file/0015/113640/landline-broadband.pdf}.
\newblock 2019.

\bibitem{datadrift-8}
{RM French}.
\newblock Catastrophic forgetting in connectionist networks.
\newblock In {\em Trends in cognitive sciences}, 1999.

\bibitem{robbins1952some}
H.~Robbins.
\newblock Some aspects of the sequential design of experiments.
\newblock {\em Bulletin of the American Mathematical Society}, 58(5), 1952.

\bibitem{datadrift-a}
{Ronald Kemker, Marc McClure, Angelina Abitino, Tyler L. Hayes, and Christopher
  Kanan}.
\newblock Measuring catastrophic forgetting in neural networks.
\newblock In {\em AAAI}, 2018.

\bibitem{fair-1}
H.~F. Scheduler.
\newblock
  {https://hadoop.apache.org/docs/r2.4.1/hadoop-yarn/hadoop-yarn-site/FairScheduler.html}.

\bibitem{getmobile}
{Shadi Noghabi, Landon Cox, Sharad Agarwal, Ganesh Ananthanarayanan}.
\newblock The emerging landscape of edge-computing.
\newblock In {\em ACM SIGMOBILE GetMobile}, 2020.

\bibitem{DBLP:conf/cvpr/ShenHPK17}
H.~Shen, S.~Han, M.~Philipose, and A.~Krishnamurthy.
\newblock Fast video classification via adaptive cascading of deep models.
\newblock In {\em CVPR}, 2017.

\bibitem{compressiondrift-11}
{Shivangi Srivastava, Maxim Berman, Matthew B. Blaschko, Devis Tuia }.
\newblock Adaptive compression-based lifelong learning.
\newblock In {\em BMVC}, 2019.

\bibitem{edgevideo-2}
{Si Young Jang, Yoonhyung Lee, Byoungheon Shin, Dongman Lee, Dionisio Vendrell
  Jacinto }.
\newblock Application-aware iot camera virtualization for video analytics edge
  computing.
\newblock In {\em ACM/IEEE SEC}, 2018.

\bibitem{DBLP:conf/nips/SnoekLA12}
J.~Snoek, H.~Larochelle, and R.~P. Adams.
\newblock Practical bayesian optimization of machine learning algorithms.
\newblock In {\em NIPS}, 2012.

\bibitem{compression-4}
{Song Han, Huizi Mao, William J. Dally }.
\newblock Accelerating very deep convolutional networks for classification and
  detection.
\newblock In {\em ICLR}, 2017.

\bibitem{sweden-data}
{Sweden Data Collection \& Processing}.

\bibitem{Swersky_scalablebayesian}
K.~Swersky, R.~Kiros, N.~Satish, N.~Sundaram, M.~M.~A. Patwary, and R.~P.
  Adams.
\newblock {Scalable Bayesian Optimization Using Deep Neural Networks}.
\newblock In {\em ICML}, 2015.

\bibitem{icarl-14}
{Sylvestre-Alvise Rebuffi, Alexander Kolesnikov, Georg Sperl, Christoph H.
  Lampert}.
\newblock icarl: Incremental classifier and representation learning.
\newblock In {\em CVPR}, 2017.

\bibitem{tf-checkpoint}
{TensorFlow Training checkpoints}.
\newblock {https://www.tensorflow.org/guide/checkpoint}.

\bibitem{ion-blog}
{The Future of Computing is Distributed}.
\newblock
  {https://www.datanami.com/2020/02/26/the-future-of-computing-is-distributed/},
  2020.

\bibitem{yarn}
V.~K. Vavilapalli, A.~C. Murthy, C.~Douglas, S.~Agarwal, M.~Konar, R.~Evans,
  T.~Graves, J.~Lowe, H.~Shah, S.~Seth, et~al.
\newblock Apache hadoop yarn: Yet another resource negotiator.
\newblock In {\em Proceedings of the 4th annual Symposium on Cloud Computing},
  page~5. ACM, 2013.

\bibitem{wang2019elastic}
H.~Wang, A.~Kembhavi, A.~Farhadi, A.~L. Yuille, and M.~Rastegari.
\newblock Elastic: Improving cnns with dynamic scaling policies.
\newblock In {\em Proceedings of the IEEE Conference on Computer Vision and
  Pattern Recognition}, pages 2258--2267, 2019.

\bibitem{DBLP:journals/corr/abs-1905-13260}
Y.~Wu, Y.~Chen, L.~Wang, Y.~Ye, Z.~Liu, Y.~Guo, and Y.~Fu.
\newblock Large scale incremental learning.
\newblock In {\em IEEE CVPR}, 2019.

\bibitem{distribution-20}
{Xi Yin, Xiang Yu, Kihyuk Sohn, Xiaoming Liu and Manmohan Chandraker}.
\newblock Feature transfer learning for face recognition with under-represented
  data.
\newblock In {\em IEEE CVPR}, 2019.

\bibitem{compression-5}
{Xiangyu Zhang, Jianhua Zou, Kaiming He, and Jian Sun}.
\newblock Deep compression: Compressing deep neural networks with pruning,
  trained quantization and huffman coding.
\newblock In {\em IEEE PAMI}, 2016.

\bibitem{DBLP:conf/osdi/XiaoBRSKHPPZZYZ18}
W.~Xiao, R.~Bhardwaj, R.~Ramjee, M.~Sivathanu, N.~Kwatra, Z.~Han, P.~Patel,
  X.~Peng, H.~Zhao, Q.~Zhang, F.~Yang, and L.~Zhou.
\newblock {Gandiva: Introspective Cluster Scheduling for Deep Learning}.
\newblock In {\em {USENIX} {OSDI}}, 2018.

\bibitem{incremental-13}
{Z. Li and D. Hoiem }.
\newblock Learning without forgetting.
\newblock In {\em ECCV}, 2016.

\bibitem{DBLP:conf/cloud/ZhangSOF17}
H.~Zhang, L.~Stafman, A.~Or, and M.~J. Freedman.
\newblock {SLAQ:} quality-driven scheduling for distributed machine learning.
\newblock In {\em SoCC}, 2017.

\bibitem{distribution-21}
{Ziwei Liu, Zhongqi Miao, Xiaohang Zhan, Jiayun Wang, Boqing Gong, Stella X. Yu
  }.
\newblock Large-scale long-tailed recognition in an open world.
\newblock In {\em CVPR}, 2019.

\end{thebibliography}


\end{document}